\documentclass[5p,times,twocolumn]{elsarticle}
\usepackage{graphicx}
\usepackage{txfonts}
\usepackage{epsfig}
\usepackage{amssymb}
\usepackage{amsthm}
\usepackage{amsmath}
\usepackage{subfig}
\usepackage{caption}
\usepackage{lineno,hyperref}
\usepackage{url} 
\biboptions{numbers,sort&compress}
\usepackage[english]{babel}
\journal{Nuclear Instruments and Methods in Physics Research A}
\begin{document}

\begin{frontmatter}

\title{Time resolution of time-of-flight detector based on multiple scintillation counters\\ readout by SiPMs}

\author[label1]{P.~W.~Cattaneo}
\author[label2]{M.~De~Gerone} 
\author[label2,label3]{F.~Gatti} 
\author[label7]{M.~Nishimura} 
\author[label4]{W.~Ootani}
\author[label1]{M.~Rossella} 
\author[label5]{S.~Shirabe} 
\author[label4]{Y.~Uchiyama  \corref{cor1}}
\cortext[cor1]{Corresponding author. Tel.: +81-3-3815-8384; fax: +81-3-3814-8806.}
        \ead{uchiyama@icepp.s.u-tokyo.ac.jp }

\address[label1] {INFN Sezione di Pavia, Via A. Bassi 6, 27100 Pavia, Italy}
\address[label2] {INFN Sezione di Genova, Via Dodecaneso 33, 16146 Genova, Italy}
\address[label3] {Dipartimento di Fisica, Universit\`a degli Studi di Genova, Via Dodecaneso 33, 16146 Genova, Italy}
\address[label7] {Department of Physics, The University of Tokyo, 7-3-1 Hongo, Bunkyo-ku, Tokyo 113-0033, Japan}
\address[label4] {ICEPP, The University of Tokyo, 7-3-1 Hongo, Bunkyo-ku, Tokyo 113-0033, Japan}
\address[label5] {Department of Physics, Kyushu University, 6-10-1 Hakozaki, Higashi-ku, Fukuoka 812-8581, Japan}

\begin{abstract}
% Background
A new timing detector measuring $\sim\!50~\mathrm{MeV}/c$ positrons is under development for the MEG~II experiment, aiming at
 a time resolution $\sigma_t \sim 30~\mathrm{ps}$. 
The resolution is expected to be achieved by measuring each positron time with multiple counters made of plastic scintillator readout by silicon photomultipliers (SiPMs).
% % Purpose
The purpose of this work is to demonstrate the time resolution for $\sim\!50~\mathrm{MeV}/c$ positrons using prototype counters.
% Methods
Counters with dimensions of $90\times 40\times 5~\mathrm{mm}^3$ readout by six SiPMs (three on each $40\times 5~\mathrm{mm}^2$ plane) were built with SiPMs from Hamamatsu Photonics and AdvanSiD
and tested in a positron beam at the DA$\Phi$NE Beam Test Facility. 
% Results
The time resolution was found to improve nearly as the square root of the number of counter hits.
A time resolution $\sigma_t=26.2\pm1.3~\mathrm{ps}$ was obtained with eight counters with Hamamatsu SiPMs.
% Conclusion
These results suggest that the design resolution is achievable in the MEG~II experiment. 
\end{abstract}


\end{frontmatter}

%%%%%%%%%%%%%%%%%%%%%%%%%%%%%%%%%%%%%%%%%%%%
\section{Introduction}
%SiPM-based scintillation counter
Timing detectors with time resolutions\footnote{Resolutions are always quoted as RMS deviation.}
 $\sigma_t \lesssim 100~\mathrm{ps}$ have been built and operated in the past decades; among them are time-of-flight detectors based on scintillation counters with photomultiplier tubes (PMTs) widely used in high-energy and nuclear physics experiments.
Nowadays, many developments of timing detectors aim at $\sigma_t\sim \mathcal{O}(10~\mathrm{ps})$ with a variety of new detector technologies.
Silicon photomultipliers (SiPMs) have received a lot of attention as a good replacement for PMTs also from the timing performance viewpoint.
In fact, excellent intrinsic time resolutions of SiPM-based scintillation counters, better than ever achieved with PMT-based ones, were achieved in \cite{muSR} and in our previous work \cite{tcsingle_ieee_2014}.

%Application, MEG II
Focusing on this point, we are developing a new timing detector to measure the time of $\sim\!50~\mathrm{MeV}/c$ positrons with a high time resolution ($\sigma_t \sim 30~\mathrm{ps}$) in the MEG II experiment~\cite{megup}. 
The concept of the new detector is to segment the system into $\sim\!500$ scintillation counters, each of which is readout by several SiPMs \cite{ootani-nima,metcjinst2014}. 
With this configuration, each particle\rq{}s time is measured by several counters, significantly improving the resolution with respect to that of a single counter.

%Goal of this work
In this work, we measured the time resolution achievable with multiple scintillation counters readout by SiPMs in response to $\sim\!50~\mathrm{MeV}/c$ positrons.
In the ideal case, the time resolution is expected to improve as 
\begin{align}
\sigma(N) = \frac{\sigma_\mathrm{single}}{\sqrt{N}},
\end{align}
where $N$ is the number of counters used in the measurement and $\sigma_\mathrm{single}$ is the single counter resolution.
However, multiple Coulomb scattering could affect significantly the time resolution at this momentum. 
In addition, secondary particles, such as $\delta$-rays, could influence the measurement.
To examine those effects and demonstrate the $\sigma_t \sim 30~\mathrm{ps}$ resolution, we tested the MEG II prototype counters in a beam at the DA$\Phi$NE Beam Test Facility of the INFN Laboratori Nazionali di Frascati (LNF).

\section{Experimental setup}

%%%%%%%%%%%%%%%%%%%%%%%%%%%%%%%%%%%%%%%%%%%%
\subsection{The Beam Test Facility}
\label{sec:btf}
The Beam Test Facility (BTF), located in the DA$\Phi$NE collider
complex at LNF \cite{btfcomm, btftns},
is a dedicated beam-line designed to provide a collimated beam of $\mathrm{e}^\pm$ 
in the energy range 20--750~MeV with a pulse rate of 50~Hz for detector tests.

The extracted electrons (positrons) are transported to the BTF hall, where the final section is located, and
the experimental equipment under test is positioned at the exit of the line after a 0.5-mm thick beryllium window.
The pulse duration can vary from 1 to 10 ns
and the average number of electrons (positrons) per bunch ranges from 1 to $10^{10}$
\cite{btf-epac-2006}.
In this test, the beam-line was tuned to extract 48-MeV positrons with an average multiplicity of 1.9 in a 10-ns bunch.
The maximum bunch width was chosen to maximize the identification power for the beam pileup  at a given beam multiplicity.
The beam spot is elliptical with sizes $\sigma_{x} \sim 15$ mm and
$\sigma_{y} \sim 5$ mm \cite{btfagile} (the coordinates $x$ and $y$ are defined in Fig.\ref{fig:setup}). Its divergence is $\sim 5$ mrad.
%%%%%%%%%%%%%%%%%%%%%%%%%%%%%%%%%%%%%%%%%%%%

\subsection{Test counters}
\begin{figure*}[tb]
\centering
\includegraphics[width=0.8\textwidth]{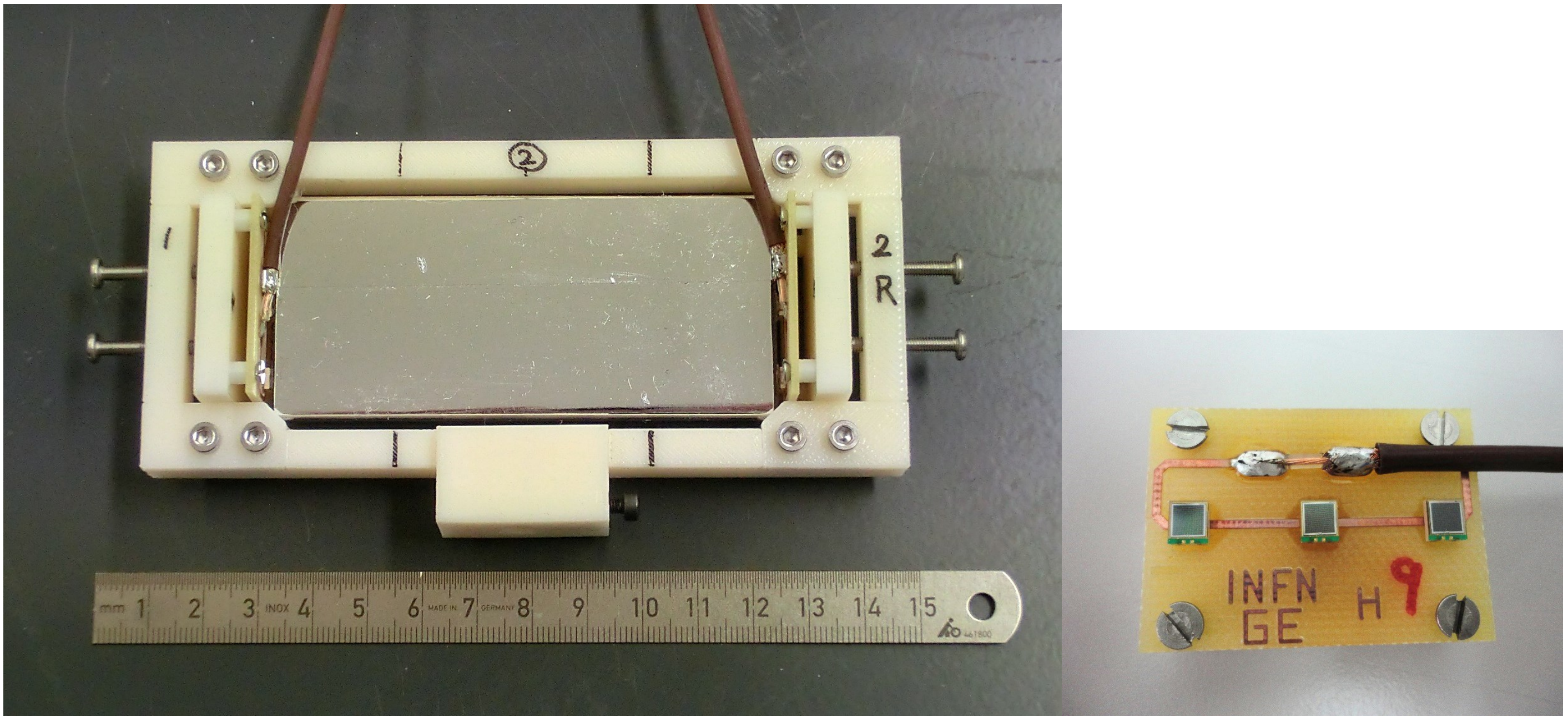}
\caption{Pictures of a test counter (left) and SiPMs (S10943-2547(X)) connected in series on a PCB (right).}
\label{fig:CounterPicture}
\end{figure*}

The design of a single counter is discussed in detail in our previous work~\cite{tcsingle_ieee_2014}.
A counter is composed of a plastic scintillator plate with dimensions of $90\times 40\times 5~\mathrm{mm^3}$ and six SiPMs. 
BC418 from Saint-Gobain Crystals is used as the scintillator material in this test (see Sec.~\ref{sec:improvement} for the discussion on the scintillator material). The scintillator is wrapped in a specular reflector (3M radiant mirror film).\footnote{A study of scintillator wrapping was carried out in \cite{ootani-nima}, in which the specular reflector was found to give the best time resolution.}
The scintillator and SiPMs are assembled in a support frame made of ABS resin as shown in Fig.~\ref{fig:CounterPicture}. Three SiPMs are optically coupled to each $40\times 5~\mathrm{mm^2}$ plane of the scintillator with optical grease (OKEN6262A).
The sensor active area is $3\times 3~\mathrm{mm}^2$ each and the fractional coverage to the scintillator cross section is 13.5\%.

%% description on the series connection
The three SiPMs on an end are connected in series and the summed signal is fed to one readout channel.  
The series connection has advantages in time measurement compared to the more conventional parallel connection because of the reduction of the total sensor capacitance.
As a consequence, the output pulse shape becomes narrower (both for the rise- and fall-times) than that from a single SiPM whereas it becomes wider with parallel connection.
Although the total charge of the single-cell-fired signal is reduced to one third of that of a single SiPM, the signal amplitude (pulse height), which is more important for time measurements, is kept comparable (compensated by the three times faster decay time) and
the signal-to-noise ratio with respect to the amplitude was measured to be 1.5 times better than that with the parallel connection.  
A direct comparison of the parallel and series connections in time measurement was carried out in \cite{nishimura_2015} where always series connection gives better time resolution at the same over-voltages $V_\mathrm{over}$.
%%%%%%%%%%%%%%%%%%%%%%%%%%%%%%%%%%%%%%%%%%%%%
%%%%%%%%%%%%%%%%%%%%%%%%%%%%%%%%%%%%%%%%%%%%

We tested two types of SiPMs.
One is S10943-2547(X) from Hamamatsu Photonics K.~K. (HPK). This is a prototype model of the commercial product S12572-050P, which introduces technologies for after-pulse suppression~\cite{hamamatsu}.
The difference is the quench resistors: this prototype model uses conventional poly-silicon quench resistors instead of metal ones.
The other is ASD-NUV-SiPM3S-P, a prototype model of NUV-SiPMs  from AdvanSiD (ASD) \cite{fbk} with 50-$\mu$m pitch micro-cells produced in early 2013.
The average HPK\rq{}s breakdown voltage is $65.1\pm0.2$~V (mean and standard deviation among 60 devices) and ASD\rq{}s one is $24.8\pm0.4$~V  (42 devices).

We define working bias-voltage ranges (in terms of $V_\mathrm{over}$) in which the dark currents are kept moderate ($\lesssim 20~\mu\mathrm{A}$). Above these ranges, the dark currents start to blow up rapidly as $V_\mathrm{over}$ increases because of the increased dark count rate (including the after-pulsing effect).
The HPK\rq{}s working range is up to $V_\mathrm{over}\sim 4$~V while that of ASD\rq{}s reaches to $V_\mathrm{over}\sim 5.5$~V.
The optimum bias voltages with respect to the time resolution for scintillation signals are given at $V_\mathrm{over}$ slightly lower than the upper limit of the working range:  $V_\mathrm{over} = 3$ and $5$~V for HPK and ASD devices, respectively (see Fig.~14 in \cite{tcsingle_ieee_2014}).  
The basic characteristics of those SiPMs were studied in \cite{tcsingle_ieee_2014},\footnote{Even though S10943-2547(X) was not used in that study, the characteristics are basically the same as those of S12572-050C(X) in \cite{tcsingle_ieee_2014}.} where
parameters relevant to time measurement were measured as functions of $V_\mathrm{over}$.
Among them, the most relevant difference between the two types is their photon detection efficiencies (PDEs): the HPK\rq{}s efficiency was measured to be about twice higher than that of ASD in the near-ultraviolet region at the optimum bias voltage for each device (refer to Fig.~10 in \cite{tcsingle_ieee_2014} for the detail).
 
We grouped three SiPMs attached to a counter end in accordance with their breakdown voltages. 
We built eight counters with HPK SiPMs and six with ASD SiPMs, 
All those were tested with $\beta$-rays from a $^{90}$Sr source ($E_\mathrm{e^-}<2.28~\mathrm{MeV}$) in advance to the beam test (see \cite{tcsingle_ieee_2014} for the setup).
The operation bias voltages applied in the beam test were determined  to optimize the counter time resolutions for the $\beta$-rays: HPK (ASD) counters were set to $V_\mathrm{over} = 3$ ($5$)~V per SiPM.\footnote{Since three SiPMs are connected in series, the applied voltages to the two terminals of a SiPM-chain are three times higher than those to single SiPM, e.g., $3\times (65.1+3.0)=204.3$~V (mean) for the HPK counters.}

\begin{figure}[t]
\centering
\includegraphics[width=3.4in]{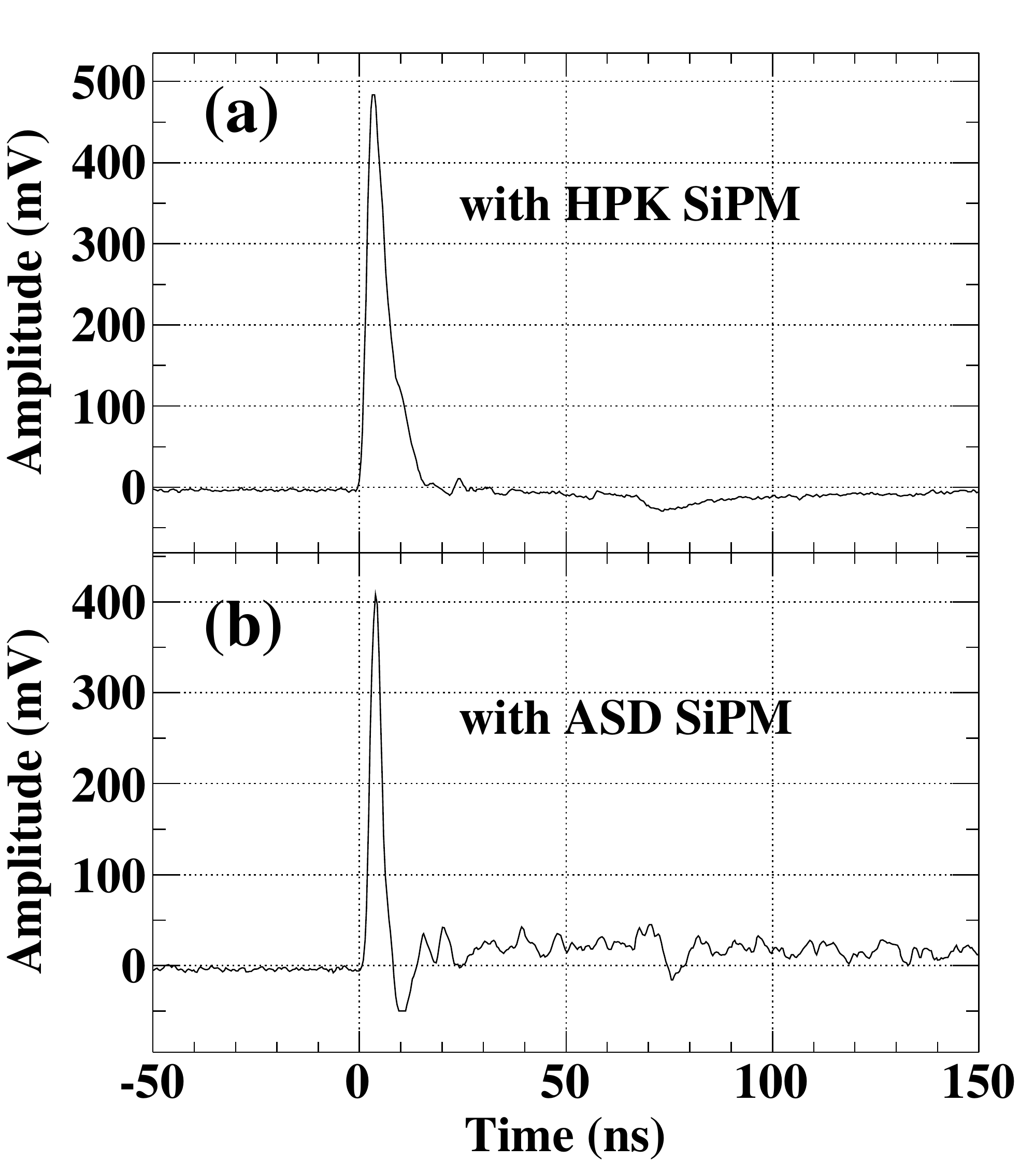}
\caption{Typical pulse shapes of scintillation signals (1 MeV energy deposition) from (a)  the HPK counters operated at $V_\mathrm{over}=3$~V and (b) the ASD ones at $5$~V. The signal from HPK counter is attenuated by a factor 2 while ASD\rq{}s is not. 
For details of the readout system see Sec.~\ref{sec:readout}. 
The bumps at around 70~ns are due to a reflection at the amplifier. 
}
\label{fig:waveform}
\end{figure}

The typical signals are shown in Fig.~\ref{fig:waveform}.
The SiPM specific long exponential tail due to the recharge current via the quench resistors is suppressed by a pole-zero cancellation circuit in the readout system to select the fast, leading-edge part of the signal and to quickly restore the baseline. For details of the readout system see Sec.~\ref{sec:readout}. 
The rise times (10\% to 90\%) of the signals are 2.2 and 1.8~ns, and the pulse widths are 5.5 and 3.3~ns at FWHM for HPK and ASD counters, respectively.
The larger fluctuation in the ASD signal tail is due to after-pulsing.
%%%%%%%%%%%%%%%%%%%%%%%%%%%%%%%%%%%%%%%%%%%%%%%
%%%%%%%%%%%%%%%%%%%%%%%%%%%%%%%%%%%%%%%%%%%%

\subsection{Beam test configuration}
\label{sec:btest}
\begin{figure*}[tb]
\centering
\includegraphics[width=0.8\textwidth]{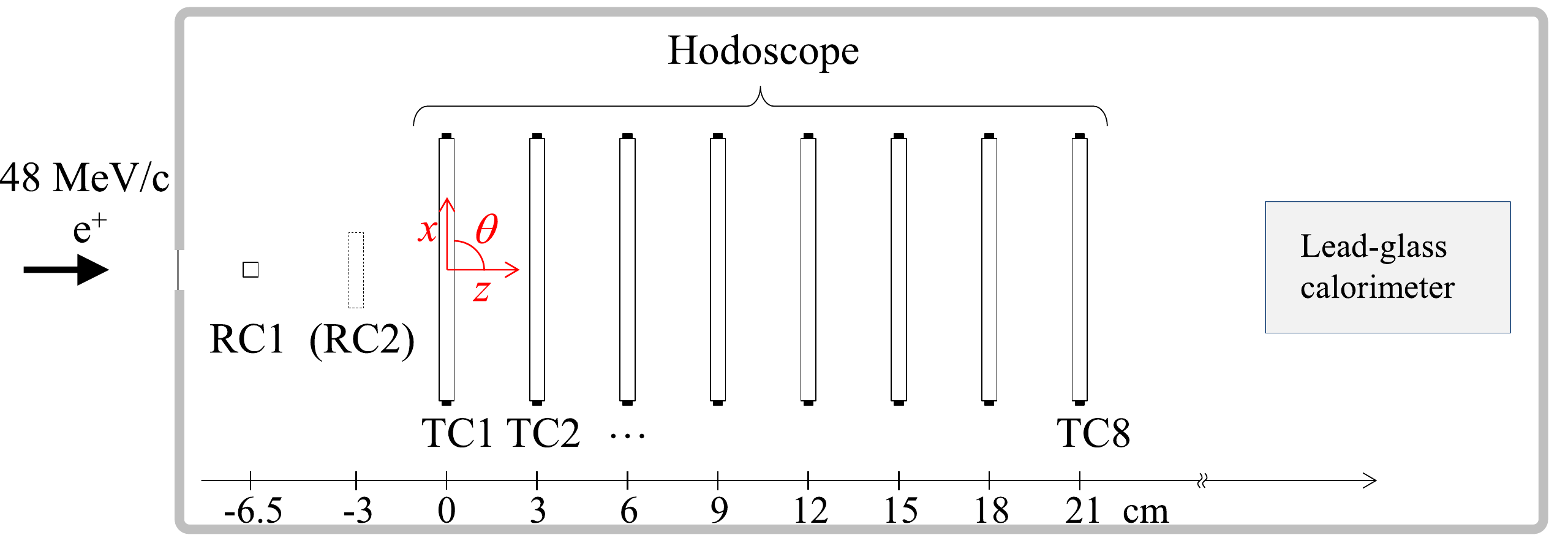}
\caption{Setup for the beam test (top view). RC$i$ and TC$i$ denote the reference counters and the test counters, respectively. See the text for the details.}
\label{fig:setup}
\end{figure*}

The setup for the beam test is schematically shown in Fig.~\ref{fig:setup}, where the coordinate system is also defined.
%%%%%%%%%%%%%%%%%%%%%%%%%%%%%%%%%%%%%%%%%%%%%
%%%%%%%%%%%%%%%%%%%%%%%%%%%%%%%%%%%%%%%%%%%%
The test counters (TC$i$, where $i$ corresponds to the position) were mounted at 3-cm intervals on a movable stage, which provides coherent rotations of all the counters as well as 2D ($x,y$) movements. 
One of the counter sets (HPK or ASD) was mounted at a time.

Additionally, two reference counters (RC1 and RC2) were prepared.
RC1 is based on $5\times 5\times 5~\mathrm{mm}^3$ BC422, wrapped in Teflon tape
and readout by an HPK SiPM S10362-33-050C.
This was always placed in front of the hodoscope for triggering and as a time reference.
RC2 is based on $25 \times 12 \times 5~\mathrm{mm}^3$ EJ232, wrapped in Teflon tape and readout by four HPK SiPMs S10362-33-050C, two connected in series at each end.\footnote{This counter was built by A.~Stoykov \cite{stoykov_NDIP}.}
This was not placed in normal data taking because of the limited number of readout channels available,
but was used in a special run, placed between RC1 and the hodoscope, to evaluate the time resolution of RC1.

A lead-glass calorimeter was located behind the hodoscope for beam monitoring.

All the counters were put inside a light-shielded box that has a thin beam entrance window made of Tedlar film and is placed so that the centers of all the counters are aligned on the beam axis.
The counter angle was normally set to $\theta=90.0^\circ$ but
we also took data with the HPK hodoscope at three other angles, $\theta=95.7^\circ$, $107.1^\circ$, and $119.2^\circ$.

The temperature in the BTF hall is controlled by an air conditioning system.
During the beam time, the temperature was kept at $24.0^\circ$--$24.5^\circ$C.

\subsection{Readout electronics and data acquisition}
\label{sec:readout}
The signal from the SiPMs' chain at each end of each counter was transmitted on a 7-m long coaxial cable (RG174, 50~$\Omega$) to an amplifier and readout by a fast sampling digitizer.
This design of the electronics chain, with SiPMs and the amplifier separated by a long cable without any pre-amplification, is convenient for many applications with space and other environmental limitations at the detector side and actually expected in the MEG II experiment.  

The amplifier is based on a two-stage voltage amplifier (MAR-6SM monolithic amplifiers from Mini-Circuits) and a pole-zero cancellation circuit with a bandwidth of 800~MHz developed at the Paul Scherrer Institut (see \cite{tcsingle_ieee_2014} for a detailed schematic).

The DRS4 evaluation board V4~\cite{drs, drsboard} was used as digitizer. 
A board has four readout channels with AC coupling at the input and an analog bandwidth of 750~MHz. 
The sampling speed was set to 2.5~GS/s. However, note that the sampling intervals are not uniform over the points \cite{drstime}; the intervals vary up to $\pm25$\% from the nominal value ($400\pm 100$~ps at 2.5~GS/s) depending on the physical sampling cells.
These intervals are constant over time and calibrated with a proper method in \cite{drstime}.
Six boards were operated in the daisy-chain mode and readout together by a front-end computer via USB.
The synchronization precision among the boards in this operation mode is, however, limited to a few hundreds ps, which is not adequate for this test.
Therefore, an external clock signal (25~MHz sine wave) was passively divided and input into the fourth channel of each board and the data on different boards are synchronized more precisely offline.

The MSCB high-voltage module\footnote{This module can supply positive voltages up to $+800$~V.} \cite{megdet} was used to supply positive bias voltages to the SiPM chains via the amplifier boards and the successive signal lines, to generate the negative polarity signal.
The signal was attenuated by a factor 2 for HPK counters (no attenuation for ASD ones) and then inverted at the DRS4 input by a transformer (ORTEC IT100, bandwidth of 440~MHz) to match its dynamic range.

A trigger was delivered by the RC1 signal alone and the trigger rate was $\sim\!4$~Hz.

%%%%%%%%%%%%%%%%%%%%%%%%%%%%%%%%%%%%%%%%%%%%
\section{Data analysis}
\subsection{Waveform analysis}
The data analysis starts with analyzing the DRS4 waveform data to extract the positron hits information.
In digital signal processing, some kinds of optimal filters can be used to recover time and amplitude information in noise. However, those are optimum only for some ideal cases, such as with well-defined constant pulse shapes of the signal and stationary white-Gaussian noises. Such requirements are not always fulfilled, in particular, not in this measurement. (Signal from the scintillator has statistical fluctuation, which is critical for timing measurement, and the noise is dominated by the SiPM dark signal, which is not white nor Gaussian.)
Another limitation comes from the characteristics of DRS4, non-equidistant sampling intervals as described in Sec.~\ref{sec:readout}, which makes it difficult to form a digital filter with keeping the best timing information.
Therefore, the use of digital filters is limited to the calculation of the pulse amplitude, charge, and baseline, for which a moving-average based digital low-pass filter with a cutoff frequency of $\sim\,120$~MHz is applied, and the timing  is calculated from the raw data.

Two time pickoff methods were tried: one is  the conventional fixed-threshold method and the other is the digital-constant-fraction (dCF) method~\cite{dcf}.
An event-by-event baseline subtraction and a cubic interpolation between the points are applied in both methods.
Since always better time resolution was obtained with the dCF method (with appropriated time-walk corrections in both cases), we adopt it in the following analysis.

The signal charge is measured by integrating the pulse for a 15-ns window from the pulse leading edge.

\subsection{Hit reconstruction}

\begin{figure}[t]
\centering
\includegraphics[width=3.5in]{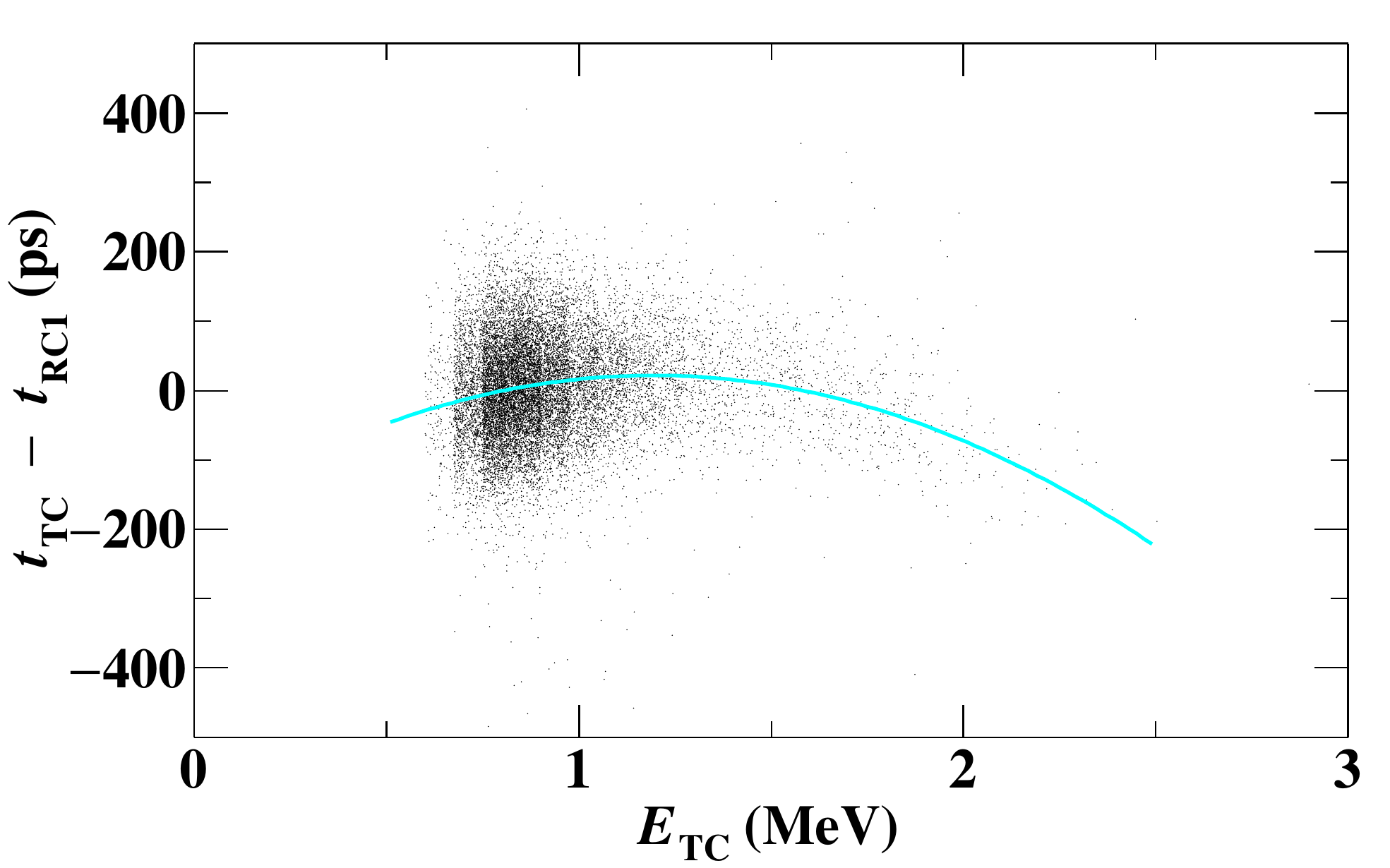}
\caption{Time walk effect as a function of energy deposited in the scintillator. The light blue curve is the best-fit quadratic function used for the correction. The event selection described in Sec.~\ref{sec:selection} is applied to make this plot.}
\label{fig:walk}
\end{figure}

The positron impact time $t_{\mathrm{TC}i}$ is computed by the average of the signal times measured at the two ends.
The residual dependence of $t_{\mathrm{TC}i}$ on the signal amplitude (time-walk effect) is observed as shown in Fig.~\ref{fig:walk}. This is caused by the nonlinear response of the amplifier and corrected for using an empirical function. 

The impact position along the long side of the counter, $l_{\mathrm{TC}i}$, is computed from the difference between the signal times at the two ends using an effective light speed, $12.0\pm0.1~\mathrm{cm/ns}$, measured using the $^{90}$Sr source.
The spatial resolution is also evaluated with the $^{90}$Sr data to be $\sigma_l = 7.5$--$8.1~(\pm 0.2)~\mathrm{mm}$, varying for counters, at 1~MeV energy deposition.

The energy deposited in a counter, $E_{\mathrm{TC}i}$, is  reconstructed by the geometric mean of the two end charges ($E_{\mathrm{TC}i}\propto(Q_{i1}Q_{i2})^{1/2}$). 
Geometric mean is chosen because the signal charges show exponential dependence on the distance between the SiPMs and the particle impact point in $l$ direction.\footnote{Since the counter size is smaller than the bulk attenuation length of the scintillator (the measured attenuation length $\lambda$ is approximately 50~cm), using arithmetic mean practically makes little difference. For scintillator with a shorter attenuation length such as BC422 ($\lambda\sim10~\mathrm{cm}$), geometric mean gives better uniformity even for this counter size.}
The energy scale is calibrated for each counter so that the most probable value of the $(Q_{i1}Q_{i2})^{1/2}$ distribution corresponds to $0.83~\mathrm{MeV}$ evaluated via a Monte Carlo (MC) simulation based on \textsc{geant4} \cite{geant4}.

%%%%%%%%%%%%%%%%%%%%%%%%%%%%%%%%%%%%%%%%%%%%%
%%%%%%%%%%%%%%%%%%%%%%%%%%%%%%%%%%%%%%%%%%%%

\subsection{Event selection}
\label{sec:selection}
A hit is defined for $E_{\mathrm{TC}i} > 0.6~\mathrm{MeV}$, fully including the Landau distribution.
To get a reliable reference time,  a more restricted selection is imposed on the energy deposited in RC1, $0.63<E_{\mathrm{RC1}}<1.10~\mathrm{MeV}$, selecting events around the Landau peak.

The number of positrons in a bunch obeys the Poisson distribution with mean 1.9 in the beam setup and hence multiple positrons sometimes impinge on the hodoscope within a bunch time ($\sim$ 10 ns):
this corresponds to an instantaneous hit rate $\sim 2/10~\mathrm{ns} = 200$~MHz.
Such a high probability of hit pileup is not expected in actual applications, being three orders of magnitude higher than that expected in the MEG II experiment.
Therefore, we select events with a single positron.

\begin{figure}[t]
\centering
\includegraphics[width=2.5in]{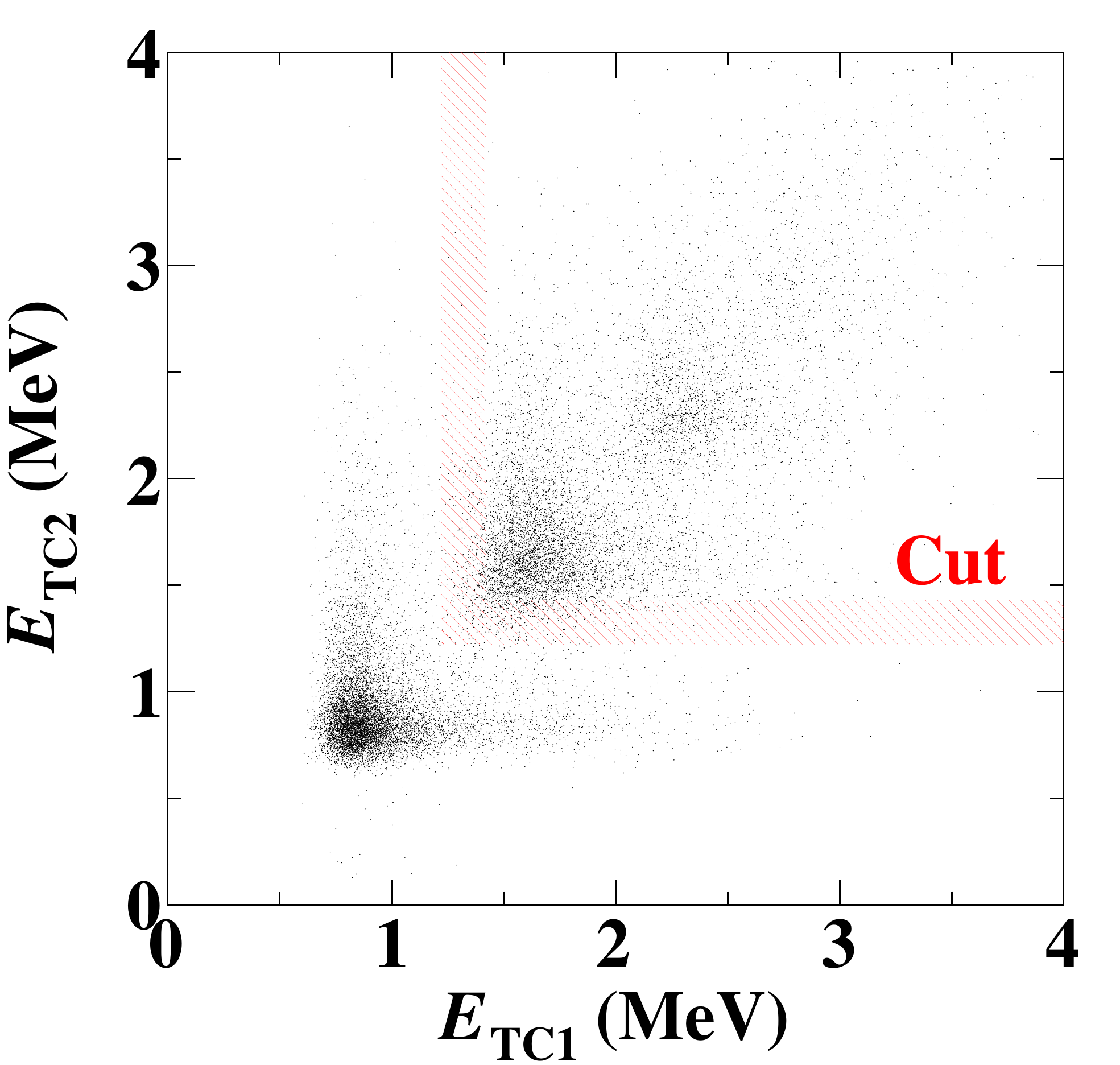}
\caption{Scatter plot of $E_\mathrm{TC1}$ vs. $E_\mathrm{TC2}$. Events in the region surrounded by the red lines are cut to reject multiple positrons events.}
\label{fig:Cut}
\end{figure}

\begin{figure}[t]
\centering
\includegraphics[width=3.5in]{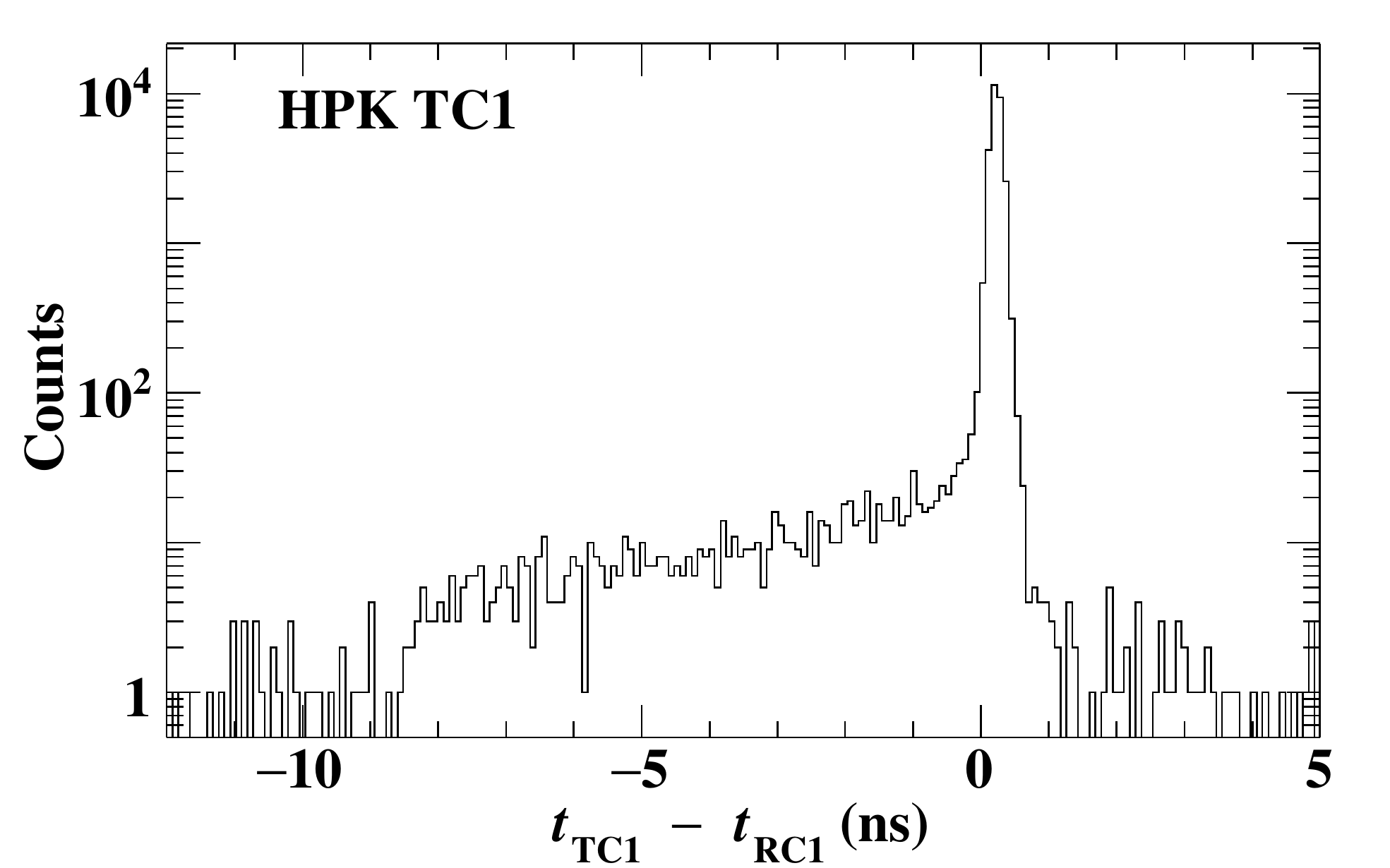}
\caption{Time distribution of TC1 relative to RC1 after the event selection criteria are imposed. The core component has $\sigma\approx80$~ps and the tail component has the beam bunch structure ($\sim\! 10$~ns), originating from pileup of beam-related background hits. Since the signal timing is subject to effect of earlier pulses, the tail distribution is asymmetric.}
\label{fig:SingleCounterTime}
\end{figure}

This is largely accomplished by imposing a selection criterion on the correlation between energies deposited in the first and second counters. 
Fig.~\ref{fig:Cut} shows the selection criterion superimposed on the scatter plot of $E_\mathrm{TC1}$ vs. $E_\mathrm{TC2}$, where the multiplicity of the beam positrons is clearly visible.
However, this criterion does not totally eliminate the overlapping hits. In particular, the hits originating from beam related photons and from scattered positrons from the support frame and other surrounding material can pass the selection criterion.
Fig.~\ref{fig:SingleCounterTime} shows the TC1 time distribution relative to the RC1 time after the selection criterion is imposed. The asymmetric tail component reflects the bunch structure of 10~ns and hence is due to the beam-related background hits.  
Against those background hits, a temporal and spatial matching among the counter hits is imposed in the data analysis, as described in Sec.~\ref{sec:clustering}. 
%%%%%%%%%%%%%%%%%%%%%%%%%%%%%%%%%%%%%%%%%%%%%
%%%%%%%%%%%%%%%%%%%%%%%%%%%%%%%%%%%%%%%%%%%%%%
%%%%%%%%%%%%%%%%%%%%%%%%%%%%%%%%%%%%%%%%%%%%%%
%%%%%%%%%%%%%%%%%%%%%%%%%%%%%%%%%%%%%%%%%%%%%%
\subsection{Track finding and fit}
\label{sec:clustering}
\subsubsection{Clustering }

%Clustering
The hit counters to be selected are different  from event to event due to the effects of the multiple scattering and the overlap of background hits.
Therefore, a clustering algorithm is applied to identify hits originating from a common particle.
It is based on requiring a rough (1~ns) time match between hits on consecutive counters while allowing one mismatched hit in between. 
The probability of getting at least one mismatched hit is 9\%.
The preliminary cluster is then passed to the tracking code.
The track finding efficiency, defined as the probability of finding a cluster with $\geq 2$ hits in the hodoscope matching with the RC1 hit, is 99.9\%.
In this analysis, events with more than one cluster are removed.

\subsubsection{Tracking}
\begin{figure}[t]
\centering
\includegraphics[width=3.in]{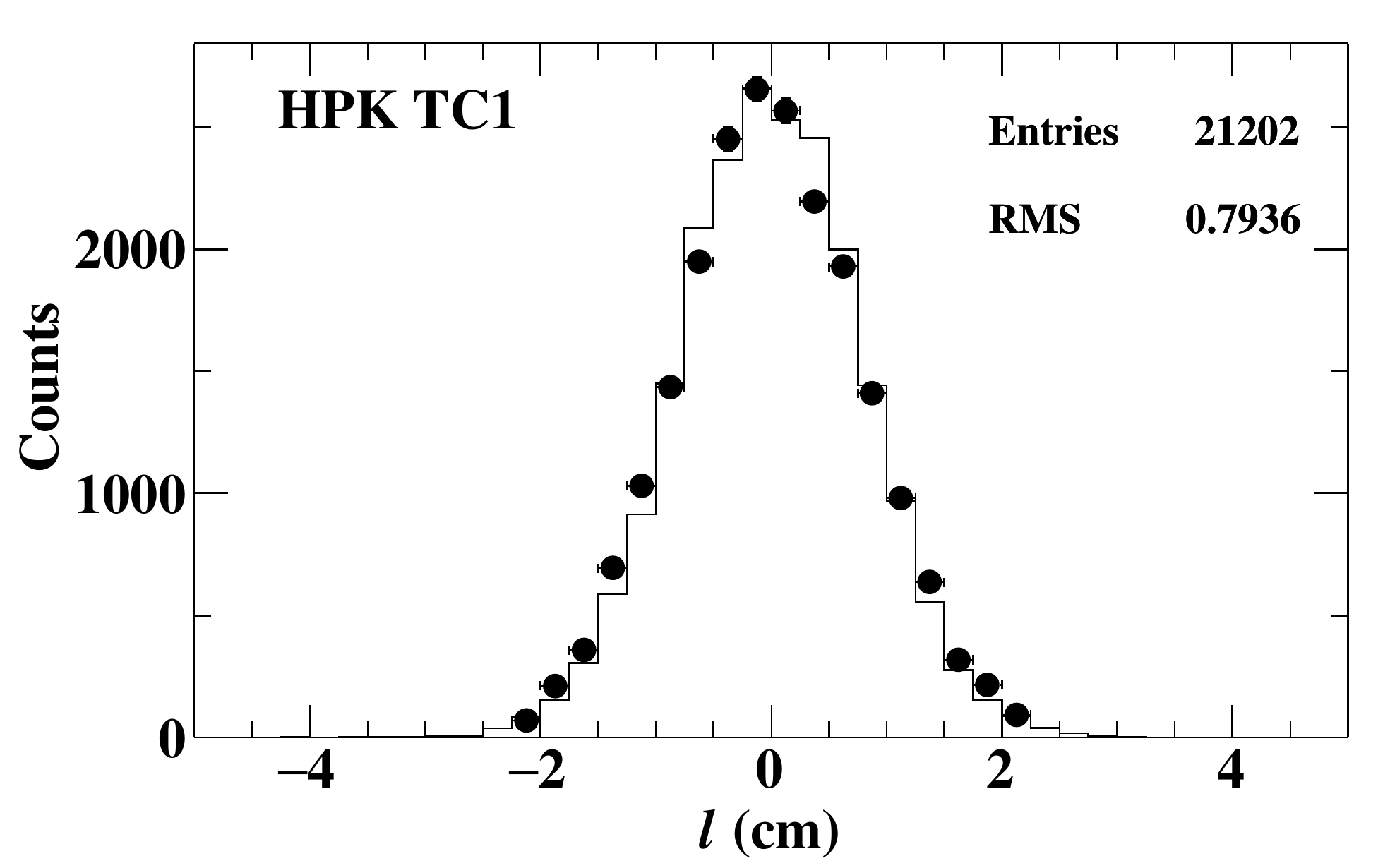}
\includegraphics[width=3.in]{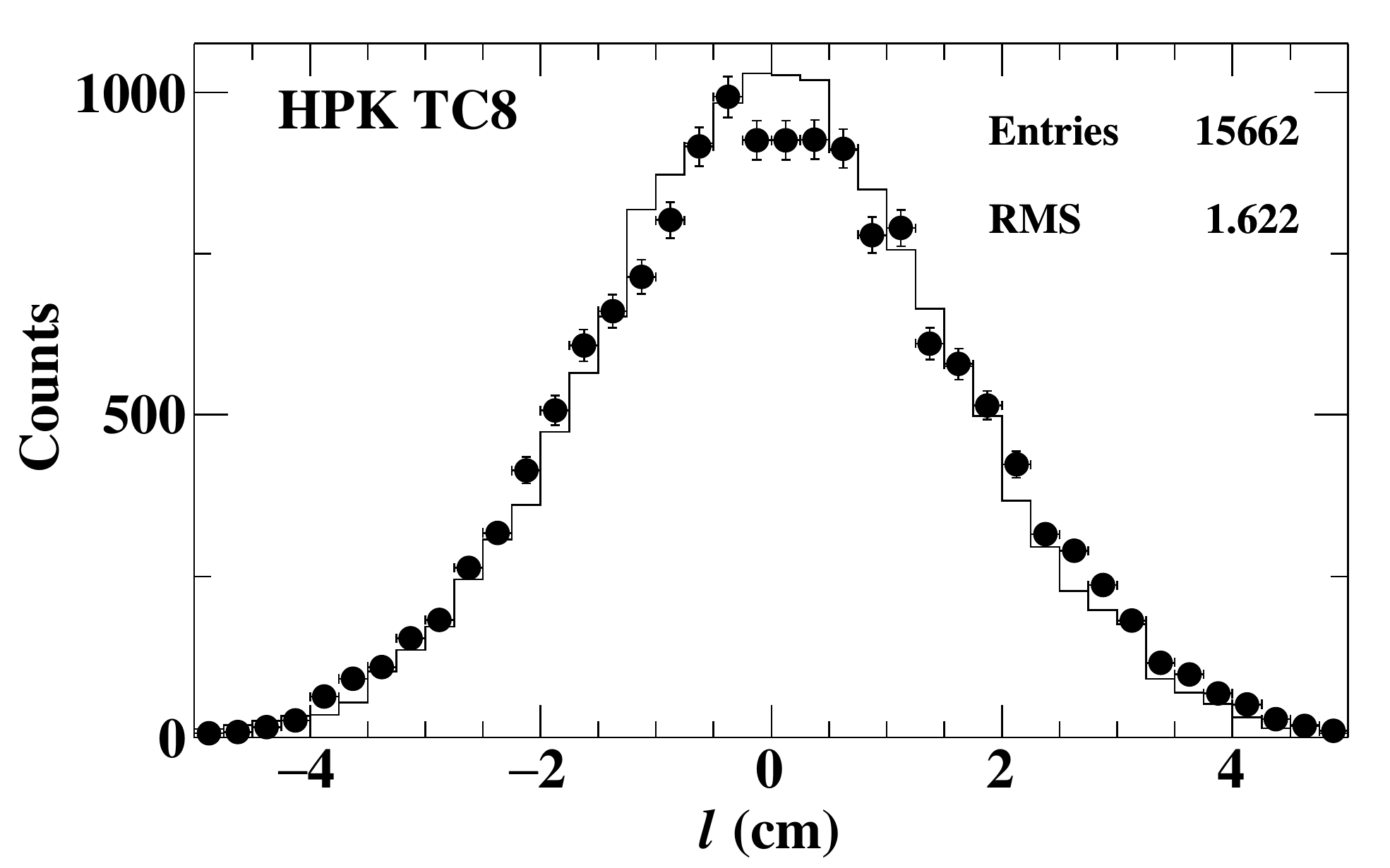}
\caption{Distributions of reconstructed hit position ($l_\mathrm{TC}$) on TC1 and TC8 ($\theta=90^\circ$). The dots are for the data and the histograms for the MC simulation normalized to the total entries in the data. Note that the widths include contribution of the position resolution ($\sigma_l\sim 8~\mathrm{mm}$).}
\label{fig:position}
\end{figure}

%Tracking
The cluster of hits provides information on the trajectory though each counter provides only one coordinate ($l$) of the impact position.
Since positrons of interest have low momenta ($\lesssim50~\mathrm{MeV}/c$), they suffer substantial multiple Coulomb scattering from the previous counters. 
A 5-mm thick scintillator plate causes a deflection of $\theta_\mathrm{MS}^\mathrm{RMS}\sim 25~\mathrm{mrad}$.
As a result, the beam is spread to $\sigma_x \sim 14~\mathrm{mm}$ at the last (eighth) counter 
as shown in Fig.~\ref{fig:position}.
Furthermore, when the counters are slanted with respect to the beam, the spread of beam causes a difference in the track length between neighbouring counters depending on the hit $l$ coordinate.
%%%%%%%%%%%%%%%%%%%%%%%%%%%%%%%%%%%%%%%%%%%
%%%%%%%%%%%%%%%%%%%%%%%%%%%%%%%%%%%%%%%%%%%%%

\begin{figure}[t]
\centering
\includegraphics[width=3.5in]{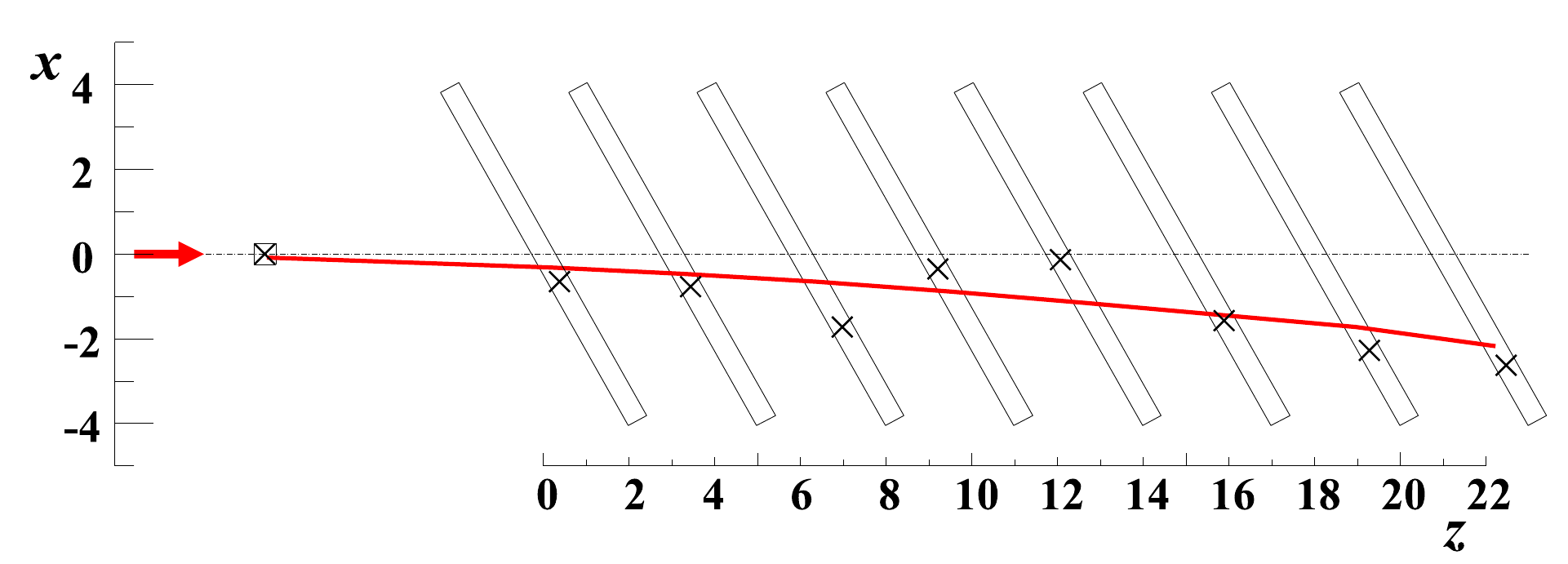}
\caption{An event display (run at $\theta=119.2^\circ$) with reconstructed hit positions (cross markers) and trajectory (red curve).}
\label{fig:tracking}
\end{figure}

To take these effects into account, a Kalman-filter based tracking algorithm \cite{Fruhwirth_1987} is applied.
Fig.~\ref{fig:tracking} shows an example of the tracking result. 
At the starting point in the forward propagation, the beam information (momentum and direction) is used.
Hits are rejected from the hit cluster if the $\chi^2$ values for the $l$-coordinate are larger than 10.
The probability of getting one or more rejected hits is 5.5\%.
The trajectory is reconstructed by connecting the fitted hit positions with line segments. 
The tracking efficiency is 99.6\%.

%%%%%%%%%%%%%%%%%%%%%%%%%%%%%%%%%%%%%%%%%%%%%%
%%%%%%%%%%%%%%%%%%%%%%%%%%%%%%%%%%%%%%%%%%%%%%%%

\subsubsection{Reconstruction of the hodoscope time}
\label{hodtime}
The time at the RC1 position is computed from the hit time of each counter ($t_{\mathrm{TC}i}$) by subtracting the time of flight from RC1 to that counter ($ t_{\mathrm{TOF}i}$) calculated from the path length.
Finally, a hodoscope time $t_\mathrm{HS}$ is derived from the average of the times measured by all hit counters:
\begin{equation*}
t_\mathrm{HS}(N) = \frac{1}{N}\sum_{i\, \in\, \mathrm{cluster}}^{} (t_{\mathrm{TC}i} - t_{\mathrm{TOF}i}), 
\end{equation*}
where $N$ is the number of hits in the cluster.

\subsection{Performance evaluation methods}
The time resolution of the hodoscope is evaluated using two methods.
One uses the difference between the time measured by the hodoscope and that by RC1:
\begin{equation*}
  \Delta t_\mathrm{RC}(N) = t_\mathrm{HS}(N) - t_\mathrm{RC1}. 
 \end{equation*}
We call this method \lq{}RC-analysis\rq{}.

The other method uses the difference between the times measured with two groups of counters (hodoscope \lq{}self-analysis\rq{}). 
For the $N$-counter resolution ($N$ is an even number), two counter groups, each of which consists of $n=N/2$ counters exclusive each other, are formed and a hodoscope time of each group $t_\mathrm{group}$ is independently calculated using the $n$ counters belonging to the group:
\begin{equation*}
t_\mathrm{group}(n) = \frac{1}{n}\sum_{i\, \in\, \mathrm{group}}^{} (t_{\mathrm{TC}i} - t_{\mathrm{TOF}i}).
\end{equation*}
In particular, two ways of grouping are tested: the odd-number-counter group and the even-number-counter group (\lq{}OE-analysis\rq{}), and the front-counter group and the back-counter group (\lq{}FB-analysis\rq{}).
The time difference between the two group times
\begin{equation*}
\Delta t_{\mathrm{OE(FB)}} (N) = t_{\mathrm{odd(front)}}(n) - t_{\mathrm{even(back)}}(n)
\end{equation*}
is calculated and $\sigma/2$ of the $\Delta t_{\mathrm{OE(FB)}} (N)$ distribution is used as a measure of the $N$-counter resolution
assuming the two group measurements are independent and their resolutions are identical. 

For the single counter resolutions, $\sigma(t_{\mathrm{TC}i} - t_{\mathrm{TC}j})/\sqrt{2}$ are used as the self-analysis. In particular, the results for $j=1$  are used as measures of the resolution of TC$i$ to cross check the RC-analysis results.

\subsubsection{Evaluation of the reference counter}
\label{sec:RC}
The precision of the reference time has to be evaluated and subtracted in the RC-analysis.
It consists of the time resolution of RC1 and the synchronization precision between the board for RC1 and those for the hodoscope.
The RC1 resolution was evaluated as follows.
First, the RC-analysis was performed with the special run in which RC2 was placed, without using it.
Then, the RC-analysis was repeated using $(t_\mathrm{RC1}+t_\mathrm{RC2})/2$ as the reference time.
The resolution of $(t_\mathrm{RC1}+t_\mathrm{RC2})/2$ was evaluated from $(t_\mathrm{RC1}-t_\mathrm{RC2})/2$.
Finally, by comparing the two RC-analyses results, the resolution of RC1 was extracted and evaluated to be $\sigma_\mathrm{RC1} = 30.3\pm 1.0~\mathrm{ps}$.
The result was cross-checked in the normal run with the same method using TC1 instead of RC2.

\subsubsection{Contribution from electronics}
The precision of synchronizing different DRS4 channels is evaluated using a pulse signal (from Phillips Scientific NIM Pocket Pulser 417) passively divided into four channels.
The jitter between two channels was measured to be $\sigma_\mathrm{sync}=12.3\pm 0.5~\mathrm{ps}$ if the two channels are on the same boards and $\sigma_\mathrm{sync}=16.5\pm0.8~\mathrm{ps}$ if the two are on different boards.

%%%%%%%%%%%%%%%%%%%%%%%%%%%%%%%%%%%%%%%%%%%%
\section{Results}
\subsection{Single-counter resolution}
%%%%%%%%%%%%%%%%%%%%%%%%%%%%%%%%%%%%%%%%%%%
\begin{figure}[t]
\centering
\includegraphics[width=3.5in]{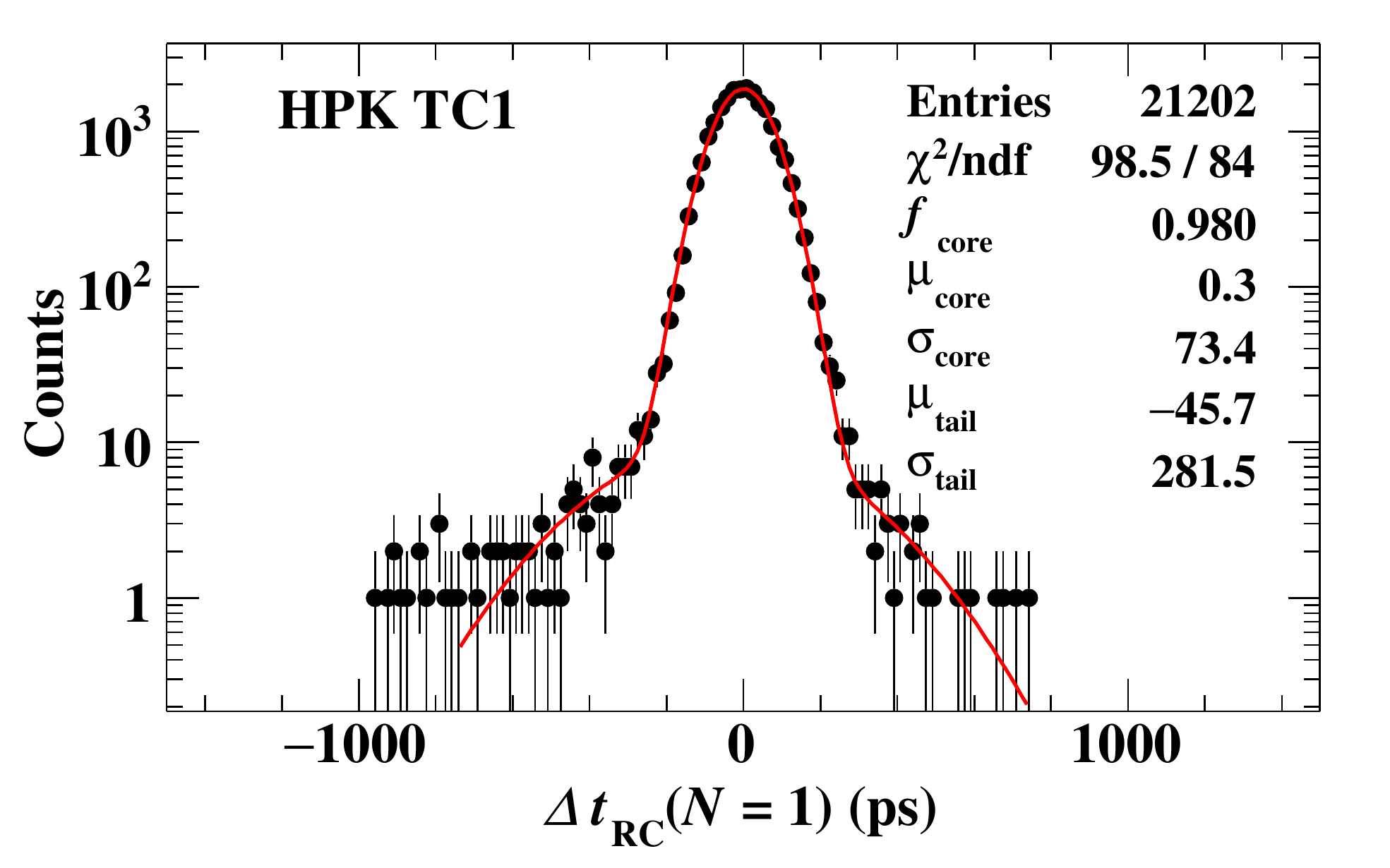}
\caption{Time distribution of single counter measurement (with HPK TC1) against RC1 time after the full reconstruction. The red curve is the sum of two Gaussian functions fitted to the data, whose parameters are shown in the legend; $f_\mathrm{core}$ is the fraction of the core component to the total, and $\mu$ and $\sigma$ are the mean and the standard deviation for each component.}
\label{fig:SingleCounterTimeAfter}
\end{figure}
\begin{figure}[t]
\centering
\includegraphics[width=3.5in]{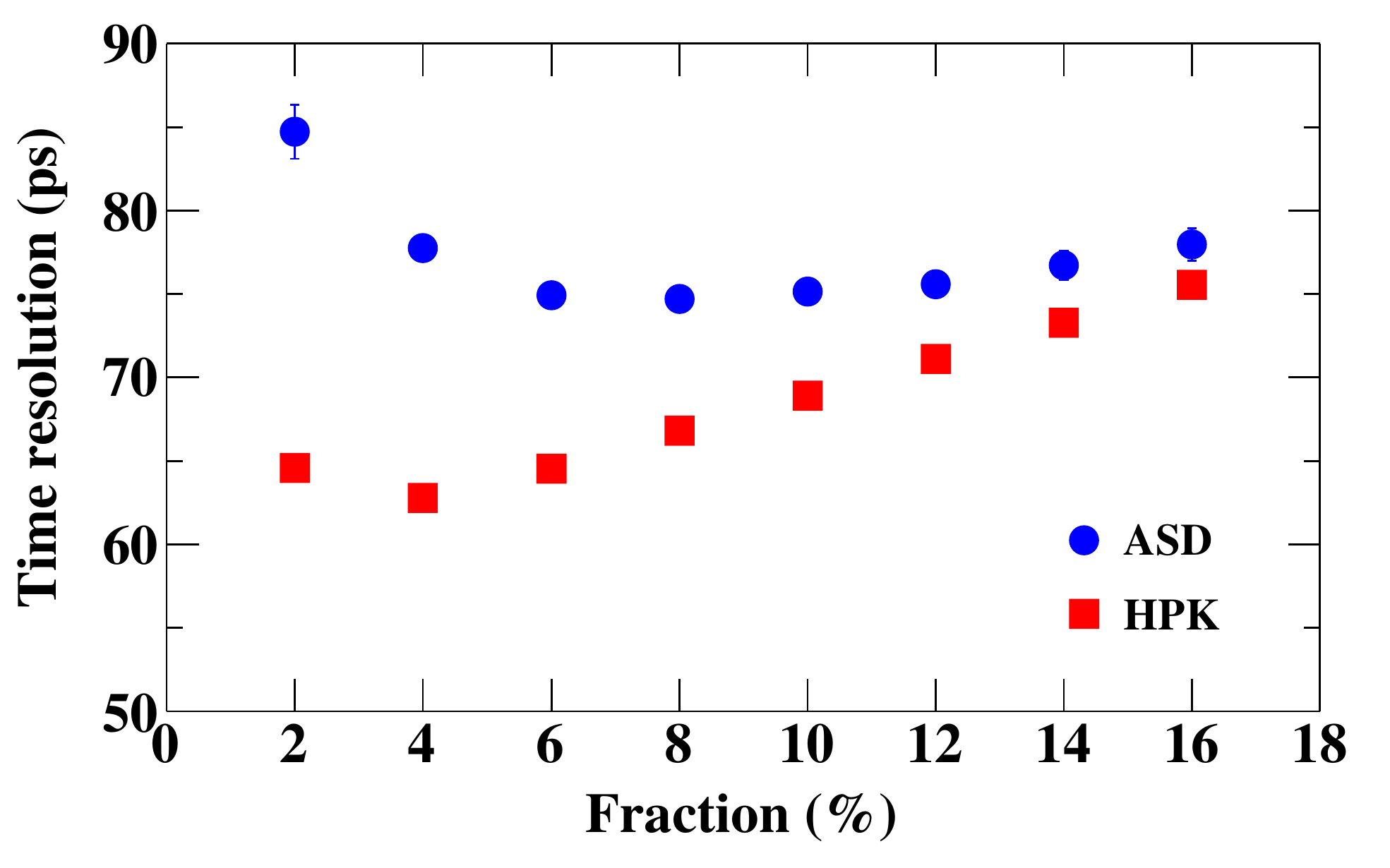}
\caption{Dependence of the average single-counter resolutions on the dCF fraction value.}
\label{fig:SingleCounterCFDependence}
\end{figure}
\begin{figure}[t]
\centering
\includegraphics[width=3.5in]{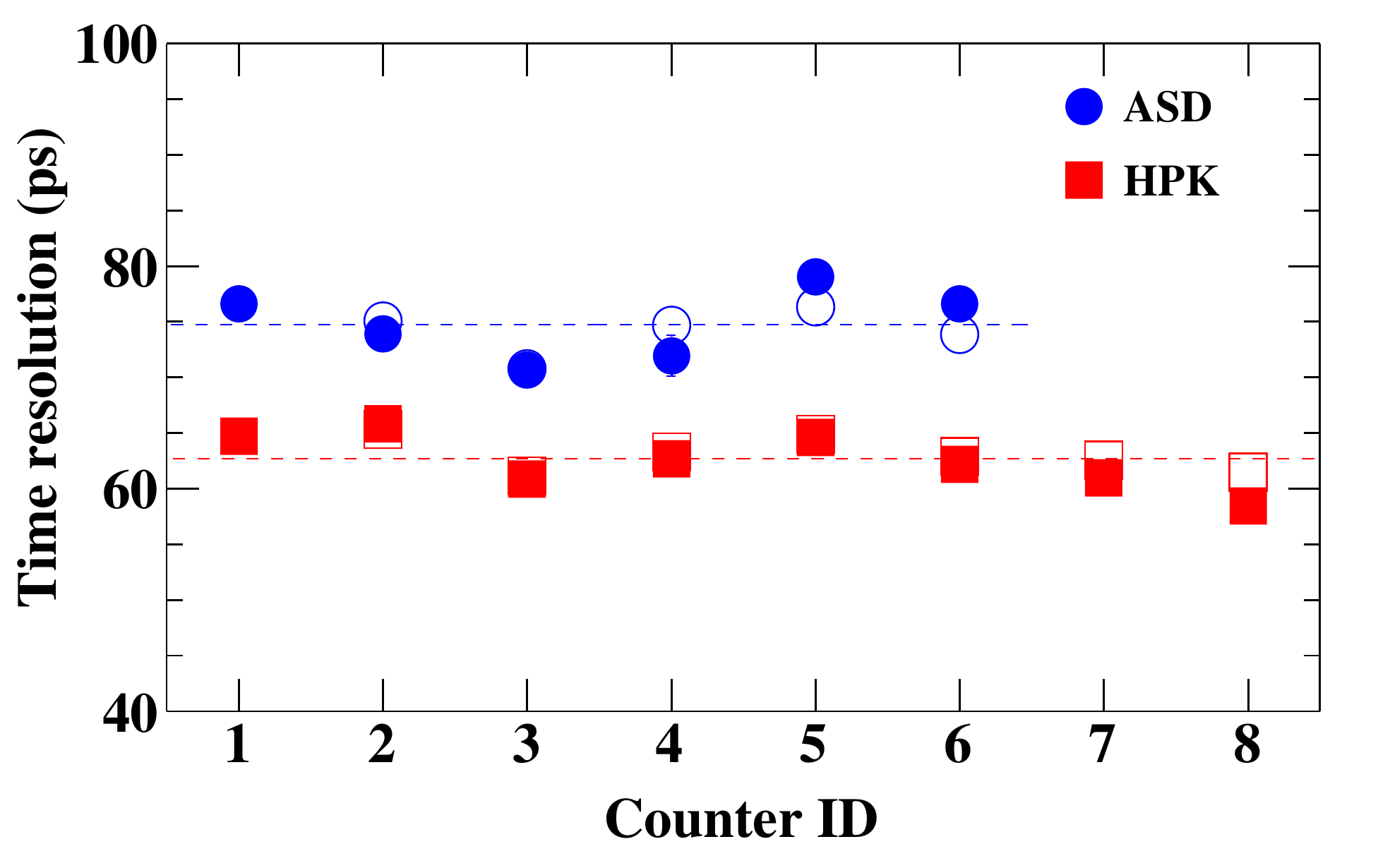}
\caption{Single-counter resolutions at the dCF fraction of 4\% (8\%) for the HPK (ASD) counters. The filled markers show the resolutions evaluated from the RC-analysis and the blank markers from the self-analysis. The statistical uncertainties are smaller than the marker size in most cases. The dashed lines show the average values in the RC-analysis.}
\label{fig:SingleCounterResolution}
\end{figure}

%%%%%%%%%%%%%%%%%%%%%%%%%%%%%%%%%%%%%%%%%%%
The typical time distribution of the single-counter measurement is shown in Fig.~\ref{fig:SingleCounterTimeAfter}.
Compared with Fig.~\ref{fig:SingleCounterTime}, the tail component due to background hits is significantly reduced by the background hits rejection in the clustering and tracking processes.
We evaluate the single-counter time resolution $\sigma_{\mathrm{TC}i}$ from the width of the core component $\sigma_\mathrm{core}$ with subtracting $\sigma_\mathrm{RC1}$ and $\sigma_\mathrm{sync}$ in quadrature. 
The dCF fraction values were scanned and the optimum values are found to be at 4\% and 8\% for HPK and ASD counters, respectively as shown in Fig.~\ref{fig:SingleCounterCFDependence}. We use these values in the following analysis.
Fig.~\ref{fig:SingleCounterResolution} shows the single-counter resolutions at the optimum dCF fractions.
The average resolutions are
%\begin{align}
\begin{equation*}
\bar{\sigma}_\mathrm{TC} = \begin{cases}
62.8\pm0.3\,\mathrm{(stat.)}\pm 0.5\,\mathrm{(sys.)}~\mathrm{ps} & (\mathrm{HPK}) \\
74.7 \pm 0.6\,\mathrm{(stat.)}\pm 0.4\,\mathrm{(sys.)}~\mathrm{ps} & (\mathrm{ASD})
\end{cases},
\end{equation*}
%\end{align}
with a standard deviations of 2.4 and 1.9~ps for HPK and ASD, respectively.

The systematic uncertainty of the single-counter resolutions predominantly comes from the uncertainty on the RC1 resolution.
The single counter resolutions are cross-checked in the self-analysis with $\sigma(t_{\mathrm{TC}i} - t_{\mathrm{TC}1})/\sqrt{2}$ and the results are superimposed in Fig.~\ref{fig:SingleCounterResolution} as the resolution of TC$i$.
The two results agree well.

\begin{figure}[t]
\centering
\includegraphics[width=3.5in]{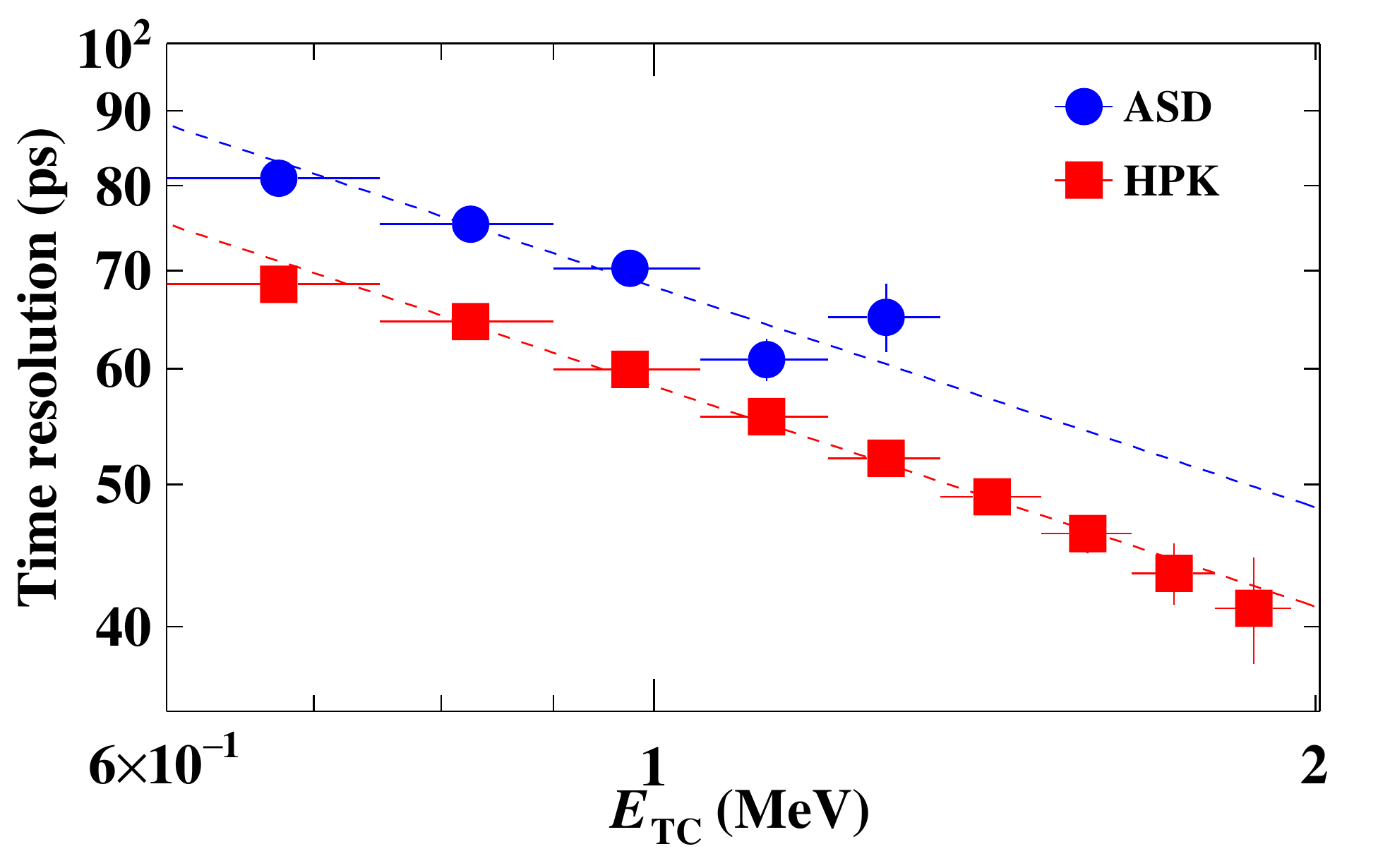}
\caption{Dependence of the single-counter resolution (average over the counters) on the deposited energy. The dashed lines are $\sigma(E_\mathrm{TC}) = \sigma^\mathrm{1MeV}/\sqrt{E_\mathrm{TC}/(1~\mathrm{MeV})}$ with the best-fit values of $\sigma^\mathrm{1MeV} = 58.5(2)$ and $68.2(6)~\mathrm{ps}$ for HPK and ASD, respectively. ASD resolution at high energies is not evaluated due to the limited statistics.}
\label{fig:SingleCounterEDependence}
\end{figure}

The resolution is expected to depend on the number of detected photons and hence on $E_\mathrm{TC}$; it is measured to be proportional to $E_\mathrm{TC}^{-1/2}$ as shown in Fig.~\ref{fig:SingleCounterEDependence}.

%%%%%%%%%%%%%%%%%%%%%%%%%%%%%%%%%%%%%%%%%%%
%%%%%%%%%%%%%%%%%%%%%%%%%%%%%%%%%%%%%%%%%%%
%%%%%%%%%%%%%%%%%%%%%%%%%%%%%%%%%%%%%%%%%%%

%%%%%%%%%%%%%%%%%%%%%%%%%%%%%%%%%%%%%%%%%%%
%%%%%%%%%%%%%%%%%%%%%%%%%%%%%%%%%%%%%%%%%%%%

\subsection{Multiple counters resolution}
\subsubsection{Overall hodoscope resolution}

\begin{figure}[t]
\centering
\includegraphics[width=3.5in]{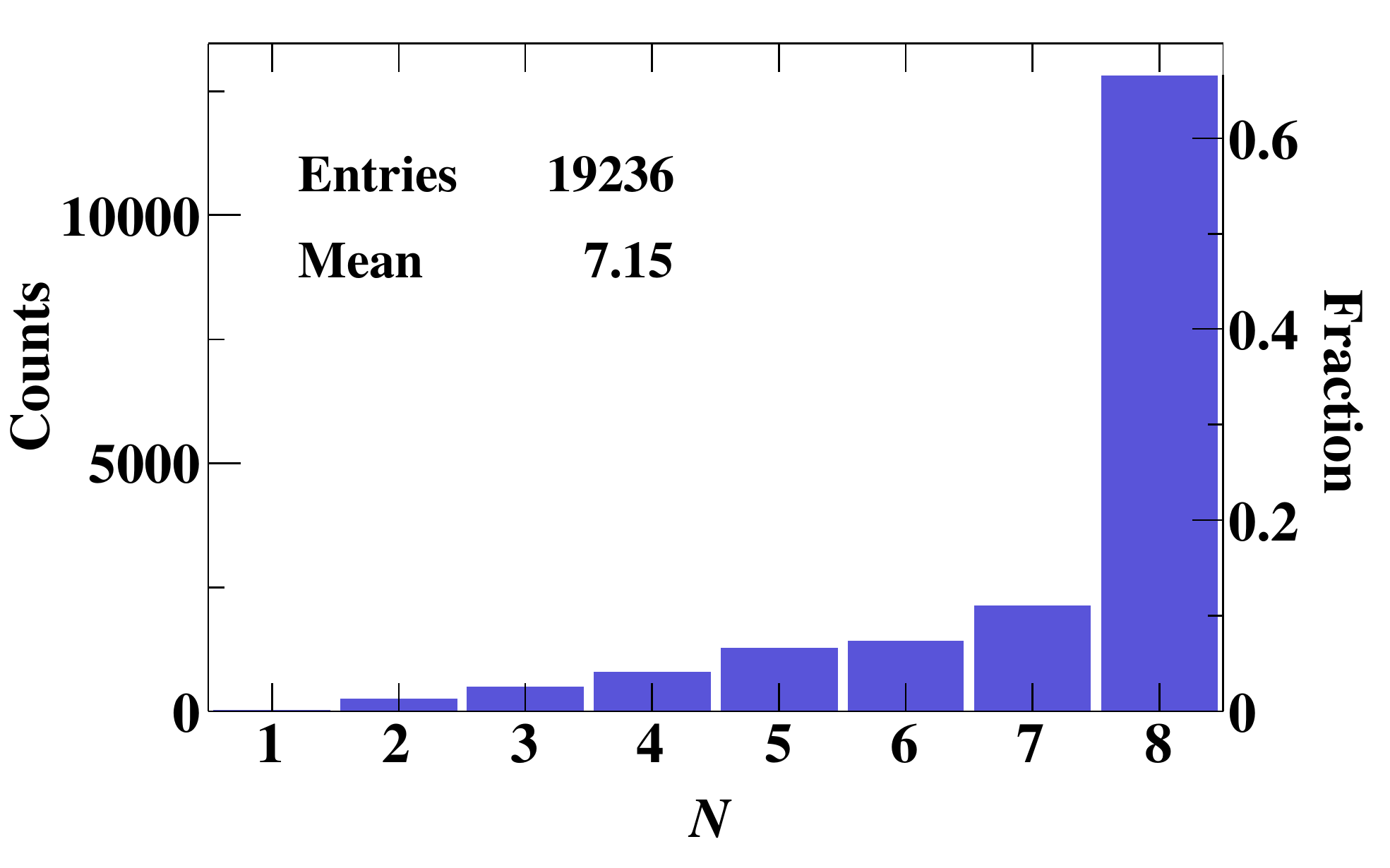}
\caption{Distribution of the number of hits in the eight-counter hodoscope.
}
\label{fig:NumberOfHits}
\end{figure}

The performance of the hodoscope depends on the distribution of $N$, which is  determined from the particle trajectories and momenta as well as the hodoscope layout and, thus, is experiment dependent.
Fig.~\ref{fig:NumberOfHits} shows the $N$ distribution in this beam test with eight HPK counters. 

\begin{figure}[t]
\centering
\includegraphics[width=3.5in]{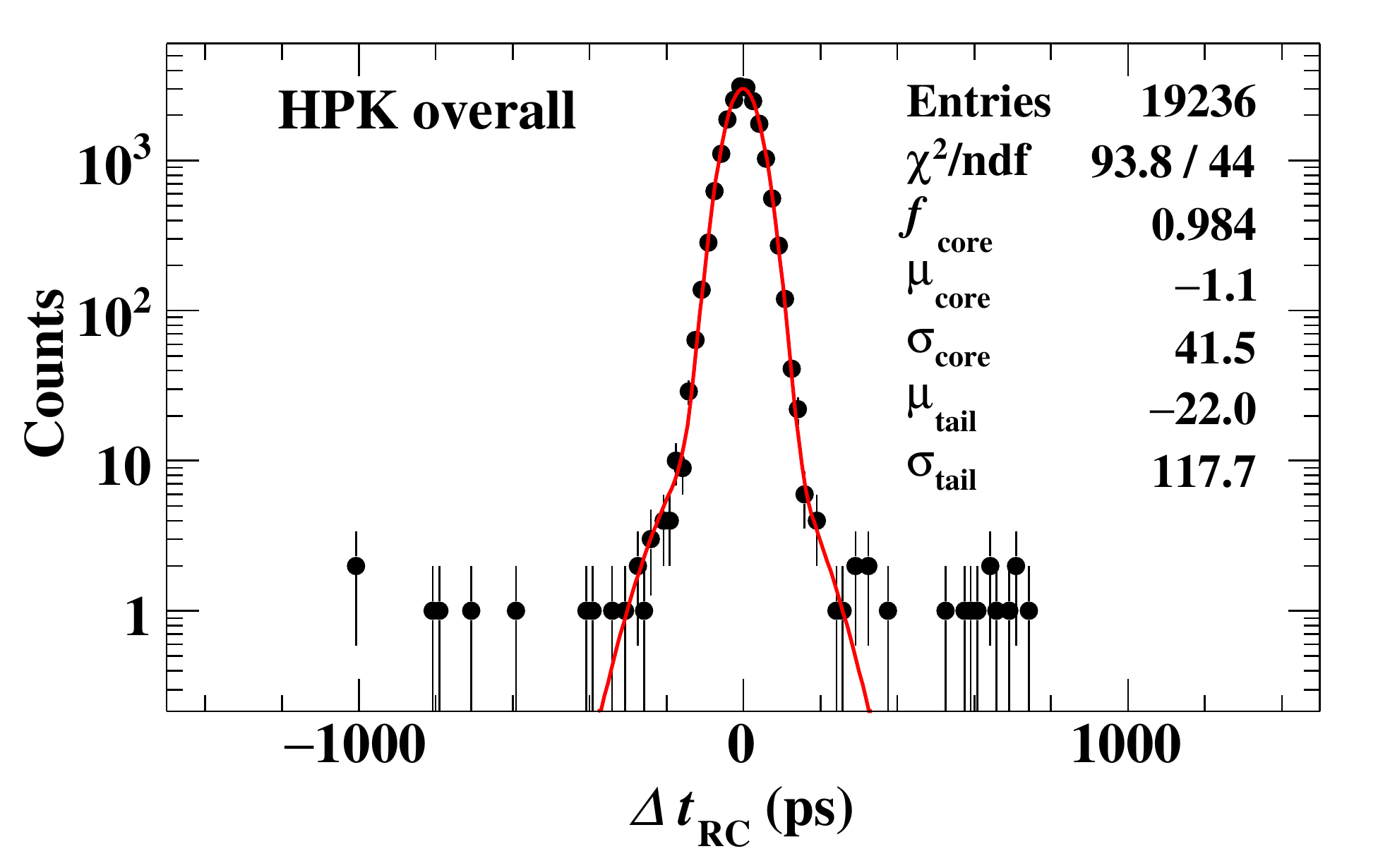}
\caption{Time distribution of the eight-HPK-counter hodoscope measurement against RC1 time for all the selected events. The fraction of the tail events, defined by $|\Delta t_\mathrm{RC}|>4\sigma_\mathrm{core}$, is 0.3\%.}
\label{fig:OverallTimeDistribution}
\end{figure}

The overall resolution of the hodoscope is evaluated with the distribution of $\Delta t_\mathrm{RC}$ accumulated over all the selected events, 
shown in Fig.~\ref{fig:OverallTimeDistribution}.
Similar to the single counter resolution, the overall resolution is extracted from the $\sigma_\mathrm{core}$ with a subtraction of the contributions from RC1 and electronics
and evaluated to be 
\begin{equation*}
\sigma_{t_\mathrm{HS}} = \begin{cases}
26.2\pm0.5\,\mathrm{(stat.)}\pm 1.2\,\mathrm{(sys.)}~\mathrm{ps} & (\mathrm{HPK}~ N_\mathrm{max} = 8) \\
33.4 \pm 1.1\,\mathrm{(stat.)}\pm 1.0\,\mathrm{(sys.)}~\mathrm{ps} & (\mathrm{ASD}~ N_\mathrm{max} = 6)
\end{cases}.
\end{equation*} 
The systematic uncertainty predominantly comes from the uncertainty on the RC1 resolution.
%%%%%%%%%%%%%%%%%%%%%%%%%%%%%%%%%%%%%%%%%%
%%%%%%%%%%%%%%%%%%%%%%%%%%%%%%%%%%%%%%%%%%%
%%%%%%%%%%%%%%%%%%%%%%%%%%%%%%%%%%%%%%%%%%%
%%%%%%%%%%%%%%%%%%%%%%%%%%%%%%%%%%%%%%%%%%%

\subsubsection{Dependence of the time resolution on $N$}
\begin{figure*}[t]
    \begin{center}
        \subfloat[RC-analysis]{
            \includegraphics[width=3.3in]{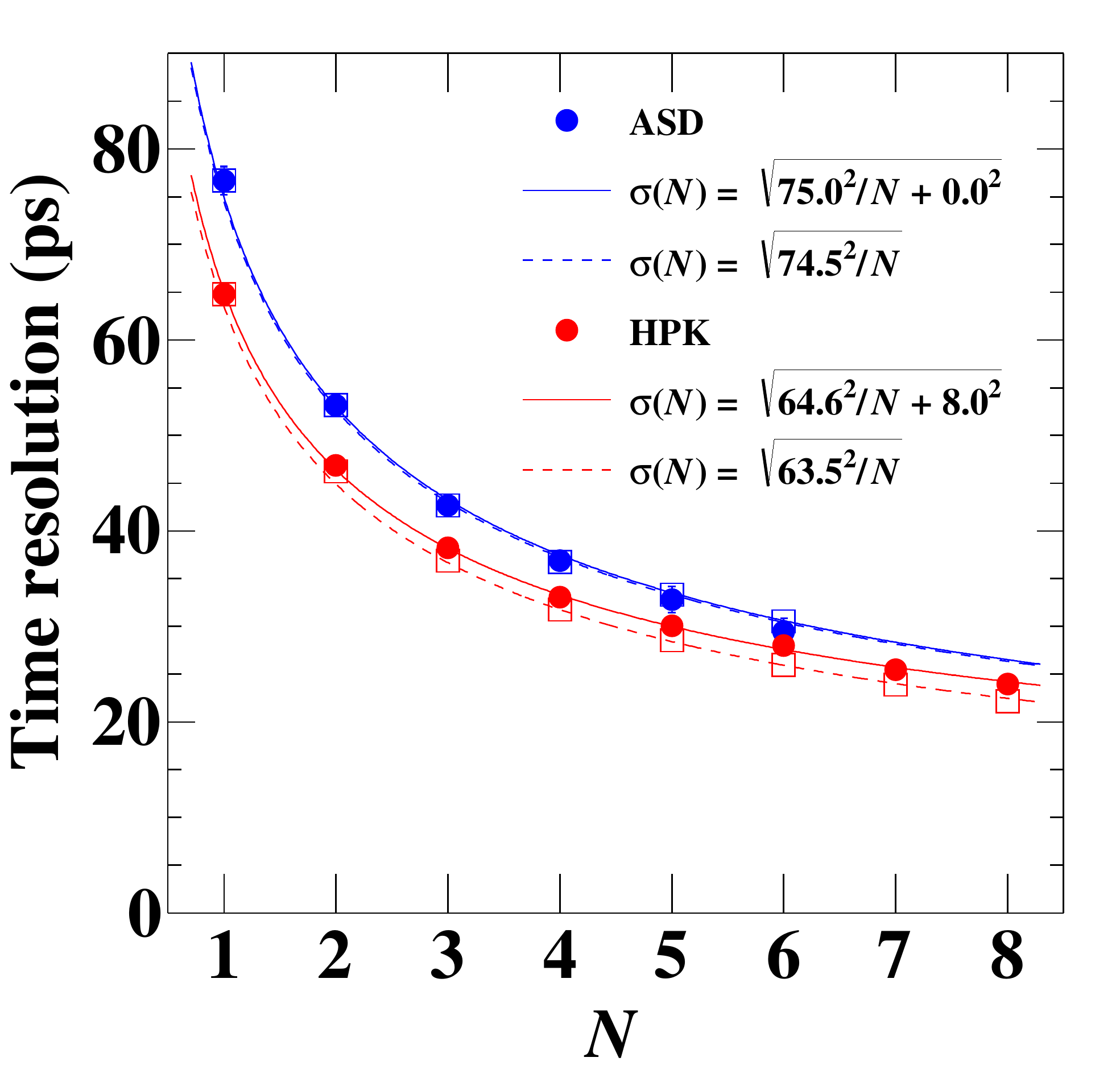}\label{fig:MultiCounterResolution}
        }
        \subfloat[OE- and FB-analyses]{
            \includegraphics[width=3.3in]{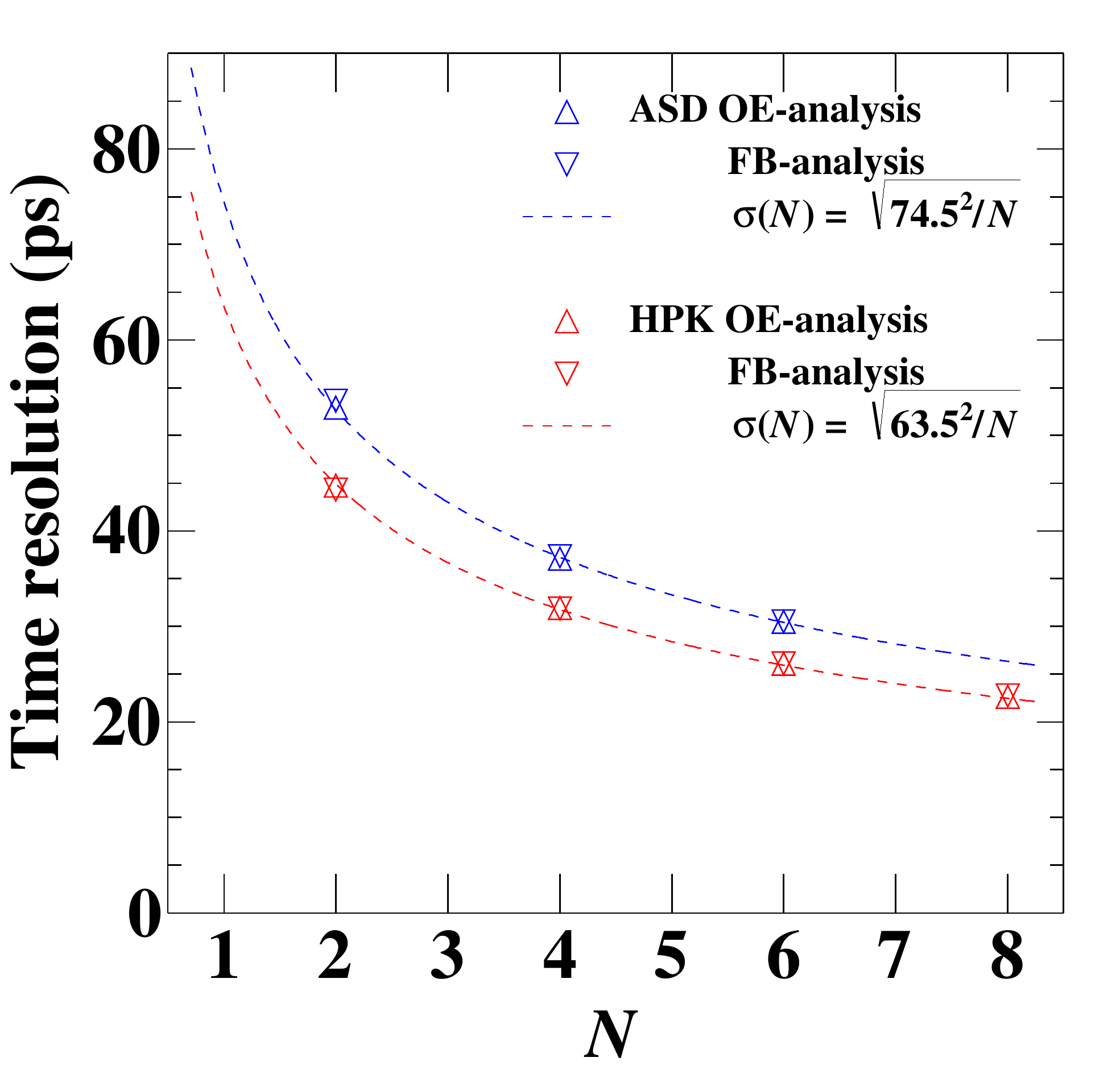}\label{fig:SelfAnalysisResolution}
        }
        \caption{Dependence of the time resolution on the number of hits. 
        \protect\subref{fig:MultiCounterResolution} The filled-circle markers show the obtained results in the RC-analysis, to which Eq.~(\ref{eq:expect2}) is fitted (solid curves). The expected behavior with the single-counter resolutions and $\sigma_\mathrm{const}=0$ is superimposed as the blank-square markers and the dashed curves.
        \protect\subref{fig:SelfAnalysisResolution} The results in the OE- and FB-analyses. The dashed curves are same as those in \protect\subref{fig:MultiCounterResolution}.}
    \end{center}    
\end{figure*}

We examined the dependence of the hodoscope time resolution on $N$. 
In this study consecutive counter hits, always starting from TC1, are used.
(An event with $k$ consecutive counter hits, is used for measurements with $N=1,2,\dots,k$.) 
The obtained hodoscope resolutions in the RC-analysis are shown in Fig.~\ref{fig:MultiCounterResolution} as a function of $N$.

%%%%%%%%%%%%%%%%%%%%%%%%%%%%%%%%%%%%%%%%%%%%%
%%%%%%%%%%%%%%%%%%%%%%%%%%%%%%%%%%%%%%%%%%%%%

The $N$-dependence of the time resolution is expected to follow
\begin{align}
\sigma(N) &= \sqrt{\frac{\sum\sigma_{\mathrm{TC}i}^2}{N^2} + \sigma_\mathrm{const}^2} \label{eq:expect1}\\
&\simeq \sqrt{\frac{\sigma_\mathrm{single}^2}{N} + \sigma_\mathrm{const}^2}.
\label{eq:expect2}
\end{align}
The first term in Eq.~(\ref{eq:expect1}) is a stochastic term coming from the counters\rq{} resolutions, represented by the parameter $\sigma_\mathrm{single}$ in Eq.~(\ref{eq:expect2}). This term decreases with $N$,
while other contributions
are parametrized by the second term $\sigma_\mathrm{const}$, which is zero in the ideal case.
The resolutions calculated from Eq.~(\ref{eq:expect1}) with the measured  $\sigma_{\mathrm{TC}i}$ and  $\sigma_\mathrm{const}=0$ are superimposed in Fig.~\ref{fig:MultiCounterResolution} for comparison.

We evaluate $\sigma_\mathrm{const}$ by fitting Eq.~(\ref{eq:expect1}) to the measured $N$-dependence (fixing $\sigma_{\mathrm{TC}i}$ to the measured values).
The best-fit values are $\sigma_\mathrm{const} = 9.3^{+3.0}_{-2.6}$ ($0.0^{+5.6}$)~ps for HPK (ASD). 
Alternatively, by fitting Eq.~(\ref{eq:expect2}) with the both parameters ($\sigma_\mathrm{single},\sigma_\mathrm{const}$) floating, we obtained the best-fit values $(\sigma_\mathrm{single},\sigma_\mathrm{const})=(64.6\pm0.6, 8.0^{+3.2}_{-2.8})$ and $(75.0\pm1.3, 0.0^{+5.0})~\mathrm{ps}$ for HPK and ASD, respectively.

Similarly, the $N$-counter resolutions are studied in the OE- and FB-analyses and the results are shown in Fig.~\ref{fig:SelfAnalysisResolution}.
In these analyses, the evaluated resolutions are consistent with the expectations with $\sigma_\mathrm{const}=0$.

\subsection{Dependence on the incident angle}
The dependence of the time resolution on the incident angle of the positrons was studied using data taken at different counter angles $\theta$.

\begin{figure}[t]
\centering
\includegraphics[width=3.3in]{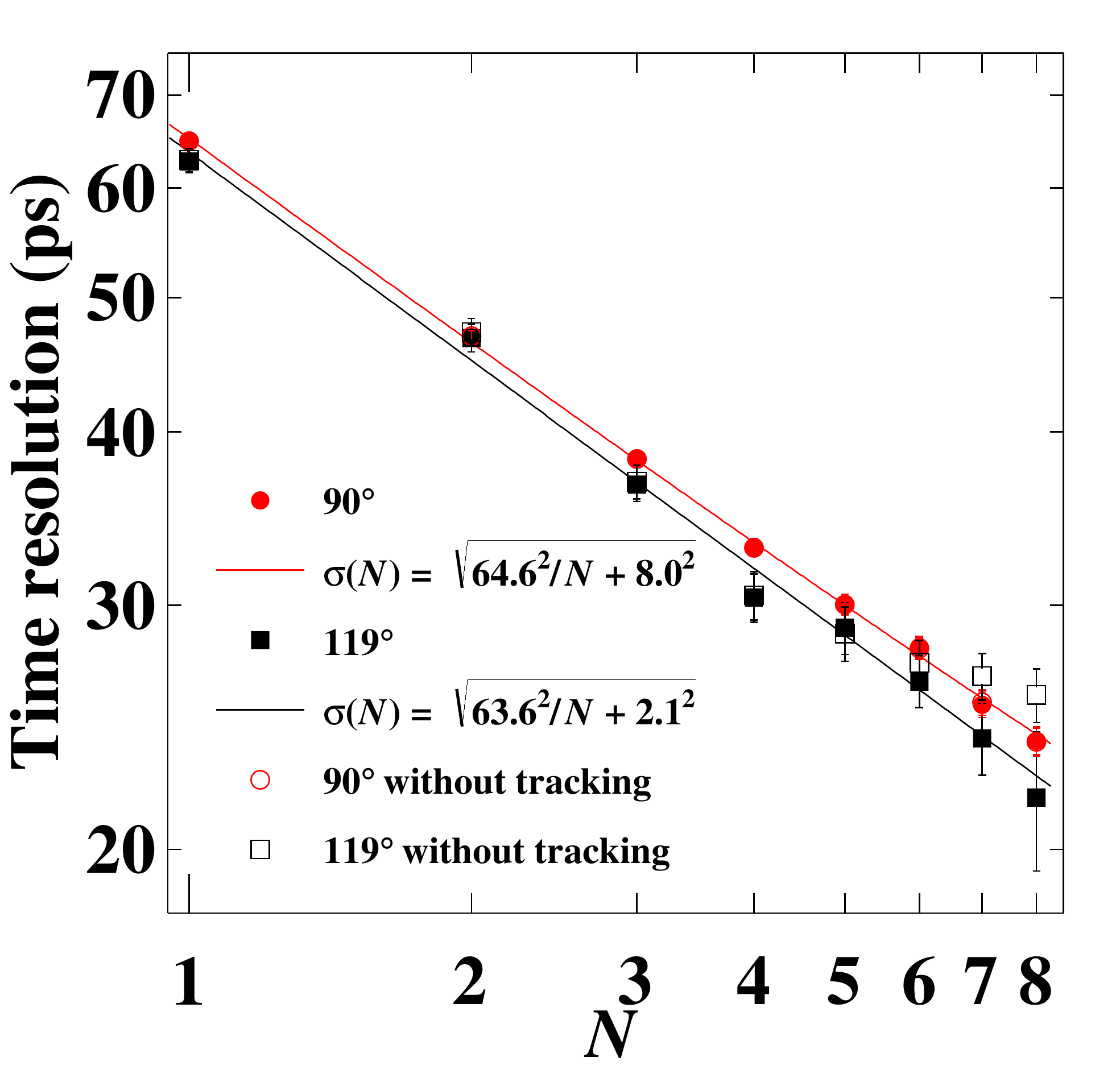}
\caption{Time resolution vs. $N$ for $\theta=90^\circ$ (red circles) and $119.2^\circ$ (black squares). The blank markers show the time resolutions calculated without the correction of time-of-flight calculated in the tracking. The effect of tracking is visible at $\theta=119.2^\circ$ and large $N$ while the results with and without tracking almost overlap at $\theta=90^\circ$.}
\label{fig:MultipleCounterResolutionAngle}
\end{figure}

If the counters are perpendicular to the beam, the variation of positron hit time due to deflection by multiple scattering is negligibly small compared to the hodoscope time resolution ($\sim 8~\mathrm{ps}$ RMS at TC8 from the MC simulation).
In contrast, when the counters are slanted, not only the deflection angle is increased by the longer effective thickness of the counters but also the track length from RC1 to TC$i$ varies with $l_{\mathrm{TC}i}$ due to the slant angle. These effects lead to larger deviations of hit times  ($\sim 28~\mathrm{ps}$ RMS at TC8). Moreover, the deviations are coherent over the counters and hence they are not diluted by averaging the measured counter times.
This effect results in a dependence of the hodoscope time resolution different from $1/\sqrt{N}$ as shown in Fig.~\ref{fig:MultipleCounterResolutionAngle} unless an appropriate correction is applied. 
However, by applying the time-of-flight correction calculated from the reconstructed trajectory, this effect can be removed and the hodoscope time resolution again follows the $1/\sqrt{N}$ dependence.
 
\begin{figure}[t]
\centering
\includegraphics[width=3.3in]{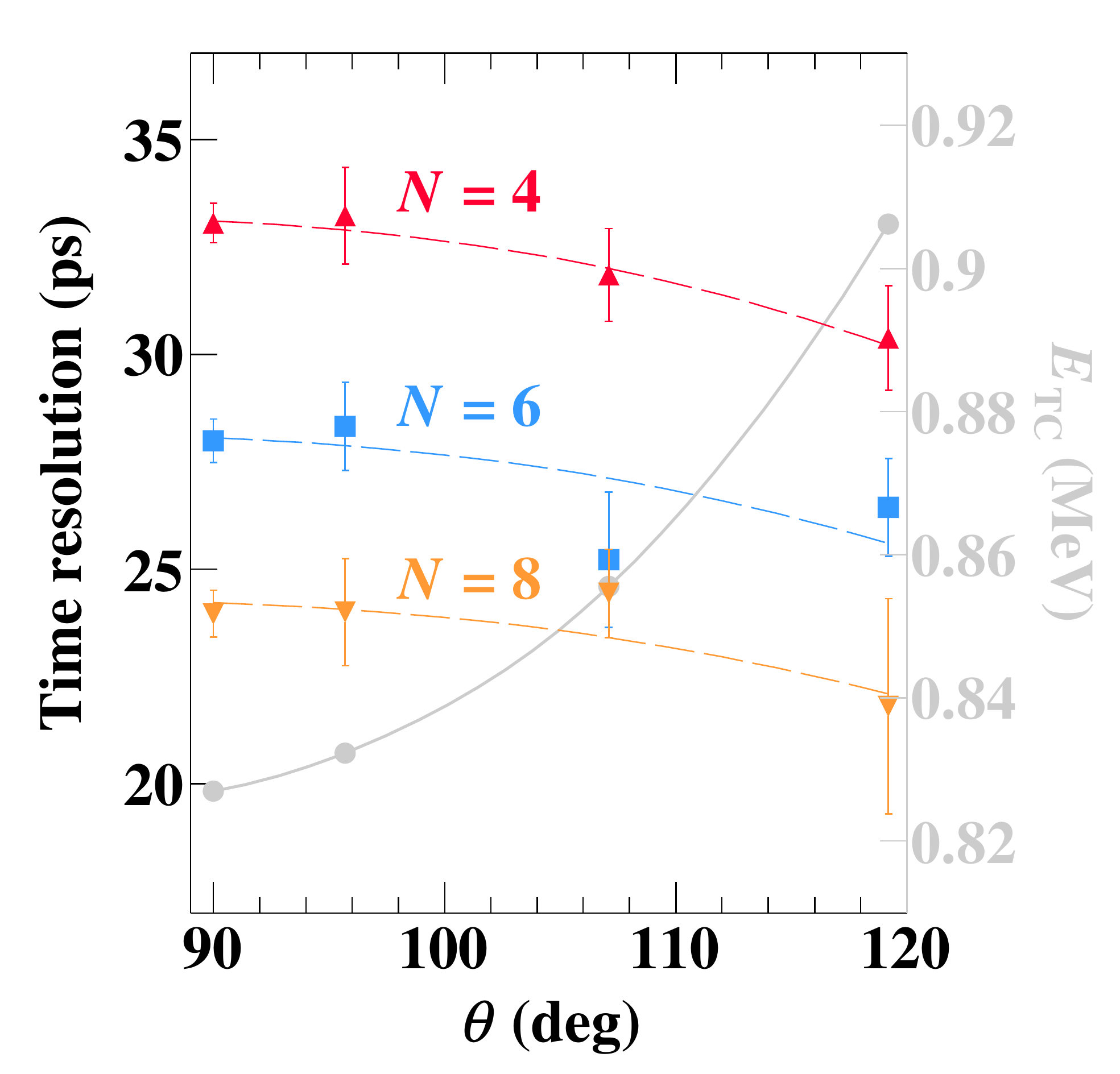}
\caption{Time resolutions ($N=4, 6, 8$) and energy deposited in the scintillator (the most probable value) vs. the counter angle $\theta$. The dashed curves are the expected behaviors by assuming $\sigma(N)\propto E_\mathrm{TC}^{-1/2}$.}
\label{fig:AngleDependence}
\end{figure}

The time resolution improves at larger incident angles. As shown in Fig.~\ref{fig:AngleDependence} this can be explained by the increase of scintillation photons due to the longer path length in the scintillator.

%%%%%%%%%%%%%%%%%%%%%%%%%%%%%%%%%%%%%%%%%%%%
%%%%%%%%%%%%%%%%%%%%%%%%%%%%%%%%%%%%%%%%%%%%
%%%%%%%%%%%%%%%%%%%%%%%%%%%%%%%%%%%%%%%%%%%%
%%%%%%%%%%%%%%%%%%%%%%%%%%%%%%%%%%%%%%%%%%%%

%%%%%%%%%%%%%%%%%%%%%%%%%%%%%%%%%%%%%%%%%%%%
\section{Discussion}

Single counter resolutions of 60--70~ps were obtained  for minimum ionizing particles.
This is comparable, for example, with PMT-based counters with similar dimensions in \cite{kim_1996}. Note that the photo-sensor coverage of our counters is limited to 13.4\% and further improvement is possible as discussed in Sec.~\ref{sec:improvement}.

More important, the time resolution was significantly improved by the multiple-counter measurement approach.
This improvement is, in principle, valid also for PMT-based counters.
In fact, multilayer configurations have sometimes been adopted to improve the resolutions, however, in most cases only with a few layers \cite{denisov_2002,besIII_2010}.
We point out that the multilayer configuration with $N\approx8$ is possible only with more compact (and also cheaper) photo-sensors such as SiPMs in actual experiments in which the counter layout would be more complicated and the available space may be limited. 
The high immunity of SiPMs to magnetic fields also extends the flexibility of the counter layout.  
We will discuss a specific application of the hodoscope to MEG II in Sec.~\ref{sec:MEG II}.

The obtained hodoscope resolution $\sigma_t=26~\mathrm{ps}$ is significantly better than those with conventional PMT-based detectors, which achieved $\sigma_t = 50$--$70~\mathrm{ps}$ \cite{sugitate_1986, kobayashi_1990, lacasse_1998, shikaze_2000, mice, tc}. 
This improvement comes at the cost of an increase in the number of readout channels whereas the total thickness of the detector is equivalent to a scintillation bar used in typical timing detectors (40~mm) \cite{tc,belleTOF_2000}. 

%The tracking capability of the hodoscope demonstrated in this test 
The tracking method was demonstrated to be effective in correctly reconstructing the path lengths between the counters in the presence of multiple scattering. 
This tracking capability of the hodoscope suggests the applicability of this technique to more complicated detector configurations without degrading the performance. Also, we point out that the tracking is useful for the identification of background hits and for a reliable interconnection between adjacent detectors such as tracker and calorimeter.

\subsection{The $\sigma_\mathrm{const}$ term}
A small but significant $\sigma_\mathrm{const}$ was observed in the RC-analysis with (and only with) the HPK hodoscope. The dominant source of the systematic uncertainty is from the uncertainty on the RC1 resolution and is already taken into account.
The nonzero $\sigma_\mathrm{const}$ term indicates positive correlations between the errors\footnote{difference between the value of a measurement and the (unobservable) true value} on $t_{\mathrm{TC}i} - t_{\mathrm{TOF}i}$ for different counters.

We investigated the origin of the term and the reason of the difference in the behavior between the HPK and ASD hodoscopes.
We find that $\sigma_\mathrm{const}$ is dependent on the dCF fraction value; $\sigma_\mathrm{const}$ negatively correlates with the fraction.
When the fraction is set to 8\% instead of 4\% for the HPK data, $\sigma_\mathrm{const}$ decreases to $5.1^{+4.6}_{-4.8}~\mathrm{ps}$, consistent with zero.
Similarly, the ASD hodoscope has a nonzero $\sigma_\mathrm{const}$ at a lower fraction: $10.5^{+3.1}_{-3.0}~\mathrm{ps}$ at a fraction 4\%.
At present, the mechanism of the dependence is not clear. However, the observed phenomena support that the origin of $\sigma_\mathrm{const}$ is in the hodoscope side such as correlation between the counters measurements, and not in the reference side. 
For this reason, the resolutions evaluated with the (OE and FB) self-analyses are subject to the systematic errors from $\sigma_\mathrm{const}$. 
Conversely, the resolutions with the RC-analysis are free from such systematic errors  (except for the associated uncertainties coming from the reference side) regardless of the unknown origin of the correlations in the hodoscope measurements. 
Therefore, we quote the RC-analysis results as the time resolutions of the hodoscope, regarded as a detector.

The optimum fraction should be determined for the best overall time resolution. 
As the fraction for the HPK hodoscope is increased from 4\%,  $\sigma_\mathrm{single}$ increases and the overall resolution is degraded. Hence, we adopt the fraction 4\% for HPK with the nonzero $\sigma_\mathrm{const}$.
The impact of $\sigma_\mathrm{const}$ on the resolution is an increase compared to the ideal case by $1.4~\mathrm{ps}$ (6\%) at $N=8$.
Therefore, though the $\sigma_\mathrm{const}$ is undesirable, the impact is marginal as far as $N\lesssim 10$.

\subsection{Application to the MEG II experiment} 
\label{sec:MEG II}
\begin{figure}[t]
\centering
\includegraphics[width=3in]{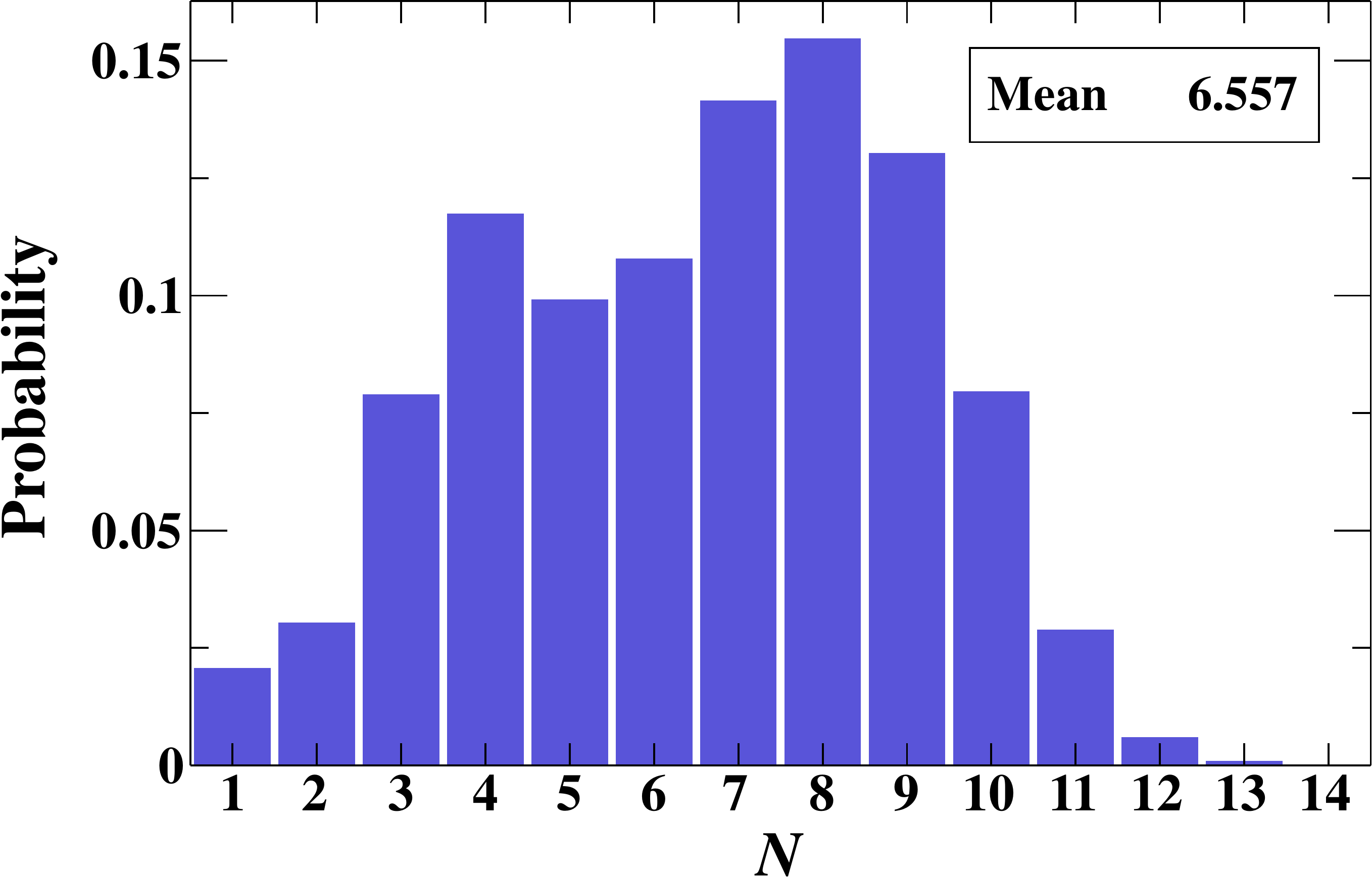}
\caption{Distribution of the number of hits from the signal positrons in the MEG II TC evaluated from a MC simulation. 
}
\label{fig:nhit_probability}
\end{figure}

Based on these results, we can estimate the time resolution expected in the MEG II Timing Counter (TC) detector.
The detector layout has been designed to
maximize $N$ while keeping the total number of counters smaller than 512 to match the limited number of readout channels
(shown in Fig.~\ref{fig:nhit_probability})  \cite{uchiyama_2013}.\footnote{In this study, we fixed the counter size to $90\times 40 \times 5~\mathrm{mm}^3$ (the same as the prototype counters in this test), while for the final design of the MEG II TC the counter size and the number of SiPMs attached to each counter will be further optimized.}
Using the expected $N$ distribution and the obtained resolutions 
(the best-fit functions of Eq.~(\ref{eq:expect2}) in the RC-analysis),
the expected time resolution, calculated by the square root of the weighted average of $\sigma^2(N)$, is estimated to be $\sigma_t = 29.6 \pm0.8$ and $33.1 \pm 0.6~\mathrm{ps}$ with HPK and ASD counters, respectively.
While HPK counters give 11\% better overall resolution even with nonzero $\sigma_\mathrm{const}$, 
both results suggest a significant improvement in comparison with MEG TC ($\sigma_t\sim 65~\mathrm{ps}$) \cite{tc} and meet our requirement.
Note that here we assume that the same precision of the tracking will be achieved in MEG II, in which tracking of positrons in an inhomogeneous magnetic field is required. 
In the MEG II configuration, a precise calculation of the trajectories requires an additional tracker (a drift chamber) before the TC.

%%%%%%%%%%%%%%%%%%%%%%%%%%%%%%%%%%%%%%%%%%%%
%%%%%%%%%%%%%%%%%%%%%%%%%%%%%%%%%%%%%%%%%%%

\subsection{Further improvements}
\label{sec:improvement}
Possible ways of performance improvement\footnote{At fixed counter size.} are two-fold: improving $\sigma_\mathrm{single}$ and increasing $N$.
Since the latter depends on the experiment layout, we only concentrate on the former possibility.

From the comparative study in \cite{tcsingle_ieee_2014}, the scintillator BC422, which has faster rise time (measured to be less than 20~ps \cite{BC422RiseTime}), is turned out to be a better choice than BC418 used in this test from the time resolution viewpoint.
The single counter resolution $\sigma_\mathrm{single} = 63~\mathrm{ps}$ (HPK) is expected to improve to $57~\mathrm{ps}$ with BC422.

It is also pointed out in \cite{tcsingle_ieee_2014} that the single counter resolution is predominantly limited by the scintillation photon statistics and can be improved by a larger fractional coverage of sensors. This can be realized by either increasing the size of each sensor or increasing the number of sensors attached to a counter.
While further study is necessary for the former option,
we successfully tested series connection up to four HPK SiPMs \cite{tcsingle_ieee_2014} and six ASD SiPMs \cite{nishimura_2014}, leading to improvement according to $\propto 1/\sqrt{N_\mathrm{SiPM}}$.

Improving the SiPM PDE is another way to increase the scintillation photon statistics.
HPK, for example, recently released a new version of their SiPMs (S13360 series) with higher fill factors and reduced cross-talk and dark count rates, resulting in 14\% higher PDE compared to one used in this test \cite{mppc13360}.
The peak sensitivity of HPK (ASD) SiPMs is at a wavelength of 450 (420)~nm \cite{mppc13360,asd_datasheet} while the peak emission from BC422 (BC418) is at 370 (391)~nm \cite{BC422_datasheet}. Therefore, further improvement would come from tuning the peak PDE to the scintillator peak emission. This could be achieved by optimizing the SiPM window material.

\section{Conclusions }
In this paper we examine the improvement in time resolution by multiple counters' measurement. 
The time resolution follows closely the expected $1/\sqrt{N}$ dependence.
We achieved a time resolution of $\sigma_t = 33.4\pm 1.5~\mathrm{ps}$ with six counters (dimensions of $90\times 40 \times 5~\mathrm{mm^3}$) readout by six ASD SiPMs (three on each  $40 \times 5~\mathrm{mm^2}$ plane) 
and $\sigma_t = 26.2\pm 1.3~\mathrm{ps}$ with eight counters  readout by six HPK SiPMs.
We conclude that a time resolution of $\sigma_t\sim 30~\mathrm{ps}$ is achievable with this technique in MEG II.

\section*{Acknowledgment}
We would like to thank the staff of BTF at INFN-LNF, in particular B.~Buonomo and P.~Valente for the arrangement of the beam.
We are also grateful to A.~Bevilacqua and F.~Siccardi of the INFN Section of Genova and F.~Barchetti and U.~Hartmann of the Paul Scherrer Institut for their valuable help.
This work was supported in part by MEXT KAKENHI Grant Numbers 22000004, 26000004.
\section*{References}
\bibliographystyle{elsarticle-my}
\bibliography{SPXBTF2013}

\end{document}